\documentclass[fleqn,usenatbib]{mnras}

\usepackage{newtxtext,newtxmath}

\usepackage[T1]{fontenc}
\usepackage{ae,aecompl}

\usepackage{graphicx}	
\usepackage{amsmath}	
\usepackage{amssymb}	
\usepackage{threeparttable}
\usepackage{longtable}
\usepackage{subcaption,booktabs}
\title[SFGs evolution at 610 MHz]{Cosmic evolution of star-forming galaxies to $z \simeq 1.8$ in the faint low-frequency radio source population}
\author[Ocran et al.]{E.\ F. Ocran$^{1,2,3}$\thanks{E-mail: ocran62@gmail.com}, A.\ R.\ Taylor$^{1,2,3}$, M.\ Vaccari$^{2,3,4}$, C.H. Ishwara-Chandra$^{3,5}$,
    \newauthor
I.\  Prandoni$^{4}$, M.\ Prescott$^{2,3}$, 
C.\  Mancuso$ ^{4}$  
\\
$^1$ Department of Astronomy, University of Cape Town, Private Bag X3, Rondebosch 7701, South Africa \\
$^2$ Department of Physics and Astronomy, University of the Western Cape, Private Bag X17, Bellville 7535, South Africa \\
$^3$ Inter-University Institute for Data Intensive Astronomy, South Africa \\
$^4$ INAF - Istituto di Radioastronomia, via Gobetti 101, 40129 Bologna, Italy\\ 
$^5$National Centre for Radio Astrophysics, Tata Institute of Fundamental Research, Pune 411007, India\\
}

\date{Accepted 2019 December 2. Received 2019 December 2; in original form 2019 September 20}

\pubyear{2019}

\usepackage{etoolbox}
\makeatletter
\patchcmd\@combinedblfloats{\box\@outputbox}{\unvbox\@outputbox}{}{\errmessage{\noexpand patch failed}}
\makeatother

\begin{document}
\label{firstpage}
\pagerange{\pageref{firstpage}--\pageref{lastpage}}
\maketitle

\parskip0pt
\baselineskip12pt

\begin{abstract}
 We study the properties of star-forming galaxies selected at  610 MHz with the GMRT in a survey covering $\sim$1.86 deg$^2$ down to a  noise of $\sim$7.1\,$\mu$Jy / beam. These were identified by combining multiple classification diagnostics: optical, X-ray, infrared and radio data.
 Of the 1685 SFGs from the GMRT sample, 496 have spectroscopic redshifts whereas 1189 have photometric redshifts. 
We find that the IRRC of
star-forming galaxies, quantified by the infrared-to-1.4 GHz radio luminosity ratio $\rm{q_{IR}}$, decreases with increasing
redshift: $\rm{q_{IR}\,=\,2.86\pm0.04(1\,+\,z)^{-0.20\pm0.02}}$ out to $z \sim 1.8$.
We use the $\rm{V/V_{max}}$ statistic to quantify the evolution of the co-moving space density of the SFG sample. Averaged over luminosity our results indicate $\rm{\langle V/V_{max} \rangle}$ to be $\rm{0.51\,\pm\, 0.06}$, which is consistent with no evolution
in overall space density. However we find $\rm V/V_{max}$ to be a function of 
radio luminosity, indicating strong luminosity evolution with redshift.
 We explore the evolution of the SFGs radio luminosity function by separating the source into five redshift bins and comparing to theoretical model predictions. 
We find a strong redshift trend that can be
fitted with a pure luminosity evolution of the form $\rm{L_{610\,MHz}\,\propto\,(\,1+\,z)^{(2.95\pm0.19)-(0.50\pm0.15)z}}$.
We calculate the cosmic SFR density since $\rm{z \sim 1.5}$ by integrating the parametric fits of the evolved 610\,MHz luminosity function. Our sample reproduces the expected steep decline in the star formation rate density since $\rm{z\,\sim\,1}$.

\end{abstract}



\section{Introduction}

 Radio continuum observations provide dust unbiased information on mechanical feedback originating in star formation and AGN radio jets
(\citealt{1992ARA&A..30..575C,2007MNRAS.381..589M,2014ARA&A..52..415M}). They thus underpin our understanding of galaxy evolution over cosmic time. Multi-wavelength analysis of the GMRT 610\,MHz deep ELAIS-N1 data down to flux densities of 50\,$\mu$Jy by \citet{2017MNRAS.468.1156O} clearly shows the transition from an AGN dominated population to a star-forming galaxy (SFG) below flux densities of $\rm{\sim\,300\mu Jy}$ 
. This is in line with what found in 1.4 GHz deep surveys.  This flux depends on the frequency.
(\citealt{1989ApJ...338...13C,2007MNRAS.375..931M,2015MNRAS.452.1263P,2018MNRAS.481.4548P}).

The synchrotron emission in SFGs is closely related to recent star formation, so that its emission is widely used as a star formation indicator. This is due to the short lifetime of the massive stars producing Type II and Type Ib supernovae (e.g. see \citealt{1992ARA&A..30..575C,2003ApJ...586..794B,2011atnf.prop.4277M}). The total infrared luminosity of a
galaxy and its total 1.4 GHz radio luminosity are known to be linearly and tightly correlated (e.g. see \citealt{1971A&A....15..110V,1985A&A...147L...6D,1985ApJ...298L...7H,1992ARA&A..30..575C,2003ApJ...586..794B,2010ApJ...714L.190S}). This so called infrared-radio correlation (IRRC) is well established for SFGs (e.g. \citealt{2010A&A...518L..31I,2010A&A...518L..28M,2014MNRAS.442..577T}).

The evolution of different radio populations conducted by using non-parametric $\rm{V/V_{max}}$ analysis (\citealt{1968ApJ...151..393S,1991ApJ...380...49M,2001ApJ...554..803Y}).
\cite{doi:10.1111/j.1365-2966.2004.07981.x} used the $\rm{V/V_{max}}$ test to show that low luminosity
radio sources evolve differently from their more powerful, predominantly Fanaroff-Riley type II (FRII).  \cite{doi:10.1111/j.1365-2966.2010.18191.x} used $\rm{V/V_{max}}$ test to investigate the cosmic evolution of low luminosity $\rm{(L_{1.4}GHz\, <\,
10^{25}W\,Hz^{−1}sr^{−1})}$ radio sources in the XMM Large Scale Structure survey field (XMM-LSS). Their results indicates that
the low luminosity sources evolve differently to their high luminosity counterparts out to a redshift of $\rm{z\sim0.8}$.
  
The bivariate luminosity function of an optical-radio matches sample describes the volume density of galaxies per unit interval of radio luminosity per interval of of optical luminosity in each redshift bin. The evolution of star-forming galaxies has been extensively  studied over the years using optical and infrared surveys. Mid and far-infrared \textit{Spitzer} observations indicate that the galaxy population undergoes pure luminosity evolution with $\rm{k_{D}\,\sim\,3.4-3.8}$ out to $\rm{z\sim1.2}$, where $\rm{k_D}$ is the pure density evolution (PDE) (e.g. \citealt{2007ApJ...660...97C,2009A&A...496...57M, 2010ApJ...718.1171R, 2011A&A...528A..35M}). While far infrared luminosity functions from Herschel data result in slightly stronger evolution estimates with $\rm{L_{\star}\,\propto\,(1\,+\,z)^{4.1\pm0.3}}$ up to $\rm{z\,\sim\,1.5}$ (\citealt{2010A&A...518L..27G, 2011ApJ...742...24L}). At low redshifts $\rm{(z\,<\,0.5)}$ Herschel studies performed by \citet{2010MNRAS.407L..69D} suggested evidence of stronger evolution in star-forming galaxies with the total luminosity density evolving as $\rm{(1 + z)^{7.1}}$ .

At radio wavelengths there has been substantial work on AGN but SFGs only become significant at low flux densities, hence are becoming more accessible with deep surveys. \cite{2007MNRAS.375..931M} studied a sample of 7824 radio sources from $\rm{1.4 \,GHz}$ NRAO Very Large Array (VLA) Sky Survey (NVSS) with galaxies brighter than $\rm{K \,=\, 12.75 \,mag }$ in the Second Incremental Data Release of 6dF Galaxy Survey (6dFGSDR2) that spanned a redshift range $\rm{0.003\,<\,z\,<\,0.3}$ and determined the local luminosity function at  $\rm{1.4 \,GHz}$ for their $60\%$ star forming galaxies (SFGs) and $40\%$ active galactic nuclei (AGN).
 \cite{2009ApJ...690..610S} derived the  cosmic star formation history (CSFH) out to $\rm{z \,= \, 1.3}$ using a sample of $\sim$350 radio selected star forming galaxies and determined an evolution in the  $\rm{1.4 \,GHz}$ luminosity function based on the VLA-COSMOS SFGs.
 \cite{2012MNRAS.426.3334M} used the Data Release 1 (DR1) from the Australia Telescope Large Area Survey (ATLAS) consisting of the preliminary data published by \cite{2006AJ....132.2409N} and \cite{2008AJ....136..519M}
et al. (2008) and reaching an rms sensitivity of
$\rm{30\, \mu\,Jy beam^{-1}}$ to derive radio luminosity functions. They constructed the radio luminosity
function for star-forming galaxies to $\rm{z\,=\,0.5}$ and for AGN to $\rm{z\,=\,0.8}$ and found that
radio luminosity function for star-forming galaxies appears to be in good agreement with previous studies.
 \cite{2013MNRAS.436.1084M} investigated the evolution of faint radio sources out to $\rm{z \, \sim\,2.5}$ by combinnig a 1 square degree  VLA radio survey complete to a depth of $100\mu Jy$ with the following surveys: Visible and Infrared Survey Telescope for Astronomy Deep Extragalactic Observations and Canadian-France-Hawaii Telescope Legacy Survey. 
\cite{2017A&A...602A...5N} use of the deep Karl G. Jansky Very Large Array (VLA) COSMOS radio observations at 3 GHz to infer radio luminosity functions of star-forming galaxies up to redshift of $\rm{z\, \sim\,5}$ based on 6040 detections with reliable optical counterparts.

In the low-frequency regime, \cite{2001MNRAS.322..536W} measure the radio luminosity function (RLF) of steep-spectrum radio sources using three redshift surveys of flux-limited samples selected at low (151 and 178 MHz) radio frequency, low-frequency source counts and the local RLF.
\citet{2016MNRAS.457..730P} presented a measurement of the evolution of SFGs to $\rm{z = 0.5}$, by matching a catalogue of radio sources measured at a frequency of 325 MHz from the Giant Metrewave Radio Telescope (GMRT) to their optical counterparts in the Galaxy And Mass Assembly (GAMA) survey. They found that the radio luminosity function at 325 MHz for SFGs closely follows that measured at 1.4 GHz.
 
 The evolution of the global galaxy SFR density can be used as a robust constraint on various simulations and semianalytic models of galaxy evolution (e.g., \citealt{1999ApJ...522..604P,2001MNRAS.320..504S, 2001ApJ...560L.131M}). Total CSFH has been constrained using MIR (24/8$\rm{\mu}$m) selected samples obtained by deep small area surveys (\citealt{2006ApJ...640..784Z,2007ApJ...660...97C,2007ApJ...663..834B}). \citet{2009ApJ...690..610S} used the VLA-COSMOS  SFGs to derive the cosmic star formation history out to $\rm{z\,=\,1.3}$. In this paper, we present a measurement of the evolution of SF galaxies to $\rm{z\,\sim\,1.5}$, by matching a catalogue of radio sources measured at a frequency of 610 MHz from the (GMRT) to their optical counterparts in the  SERVS Data Fusion\footnote{\url{http://www.mattiavaccari.net/df}} (\citealt{Vaccari2010,Vaccari2015}). The 610 MHz GMRT survey covers a sky area of $\sim$1.86 deg$^2$. The restoring beam is 6 arcsec circular and the RMS in the central region is $\sim$ 7.1\,$\mu$Jy\,beam$^{-1}$ making this survey the most sensitive low-frequency deep field to date. The SERVS Data Fusion provides reliable spectroscopic and photometric redshifts, allowing us to classify AGN and SFG. 
 The layout of this paper is as follows: we first introduce the 610 MHz SFG data in Section~\ref{sfgdata.sec}. In Section~\ref{sfgprops.sec}, we present the sample properties of the selected SFGs. The estimation of the radio luminosity function is discussed in Section~\ref{rlf.sec}. In Section~\ref{evol.sec}, we describe how we constrain the evolution of the SFG luminosity function out to $z\simeq1.5$. The implications for the cosmic star formation density are presented in Section~\ref{csfh.sec}.
 We adopt throughout the paper a flat concordance Lambda cold dark matter ($\rm{\Lambda}$CDM) ,with the following parameters: Hubble constant $\rm{H_{0}\, = \,70 \,kms^{-1}\,Mpc^{-1}}$, dark energy density $\rm{\Omega_{\Lambda}\, =\, 0.7}$ and matter density $\rm{\Omega_{m}\, =\, 0.3}$.

\section{The 610 MHz GMRT data}\label{sfgdata.sec}
The radio data we use in this paper, is taken from the GMRT at 610 MHz the covering $\sim$1.86 deg$^2$ of EN1 field. The survey consisted of 7 closely-spaced pointings. The on source integration time was $\sim$ 18 hours per pointing. The resolutions before mosaic, for each pointing were in the range 4.5 to 6 arcseconds. The minimum rms noise in the central region of the image is 7.1 $\mu$Jy\,beam$^{-1}$. The radio data is fully described in \cite{10.1093/mnras/stz2954}. Data analysis was carried  out  on  the  data  intensive cloud at the Inter-University Institute for Data Intensive  Astronomy (IDIA). A source catalogue was produced  by extracting sources in the mosaic using the \textsc{PyBDSF} source finder \citep{2015ascl.soft02007M}.
This resulted in a final catalogue of 4290 radio sources. 
By matching to multi-wavelength data  against SERVS IRAC12
positions Fusion\footnote{\url{http://www.mattiavaccari.net/df}} (\citealt{Vaccari2010,Vaccari2015}), we obtain a redshift estimate for 72\%, with 19\% based on spectroscopy. The redshift estimates  are a combination of spectroscopic and photometric redshifts from (i.e. the Hyper Suprime-Cam (HSC) Photometric Redshift Catalogue \citep{2018PASJ...70S...9T}, the revised SWIRE Photometric Redshift Catalogue \citep{RowanRobinson2013} and the  Herschel Extragalactic Legacy Project \citep[HELP]{Vaccari2016,2019MNRAS.tmp.2145S} ).
 For 3105 of the sources with redshifts we use radio and X-ray luminosity, optical line ratios, mid-infrared colors, and 24$\mu$m and IR to radio flux ratios to separate SFGs from AGN.
 In \citet{10.1093/mnras/stz2954}, we outlined that total number of sources with redshifts for which we can define at least one AGN indicator was 2305 (i.e. $\sim\,54\%$ of the whole 4290 sample and $\sim\,74\%$ of the 3105 sources with redshifts). We classified 1685 sources as SFG constituting 73\% of the 2305 sources for which we were able to define at least one AGN indicator for source classification.
For sources with redshift, rest frame 610 MHz radio luminosities are calculated using equation~\ref{lum.eqn} below:
\begin{equation}
 \centering
\rm{L_{610}\  = 4{\rm \pi} D^{2}_{L} {\frac{S_{obs}} {(1 + z) ^{1 +\alpha}}}}
\label{lum.eqn}
\end{equation}
where $\rm{L}$ is the luminosity in $\rm{W Hz^{-1}}$ at the frequency $\rm{\nu}$, $\rm{D_{L}}$ is the luminosity distance in metres. $\rm{S_{obs}}$ is the observed flux density at 610 MHz, and $\rm{\alpha}$ is the spectral index and it is defined as $\rm{S\propto\,\nu^{\alpha}}$. 
In \citet{10.1093/mnras/stz2954},  we measured a median spectral index that steepens
with frequency with
$\mathrm{\alpha^{610}_{325}\,=\,-0.80\,\pm\,0.29}$, for $\sim$479 sources and $\mathrm{\alpha^{610}_{1400}\,=\,-0.83\,\pm\,0.31}$ for $\sim$99 sources. Hence, we use  the canonical spectral index of
$\rm{\alpha\,=\,-0.8}$ often assumed for SFGs \citep{1992ARA&A..30..575C}.

\subsection{The SFG sample}\label{sfg_sub.sec}

\begin{figure}
\centering
\centerline{\includegraphics[width = 0.52\textwidth]{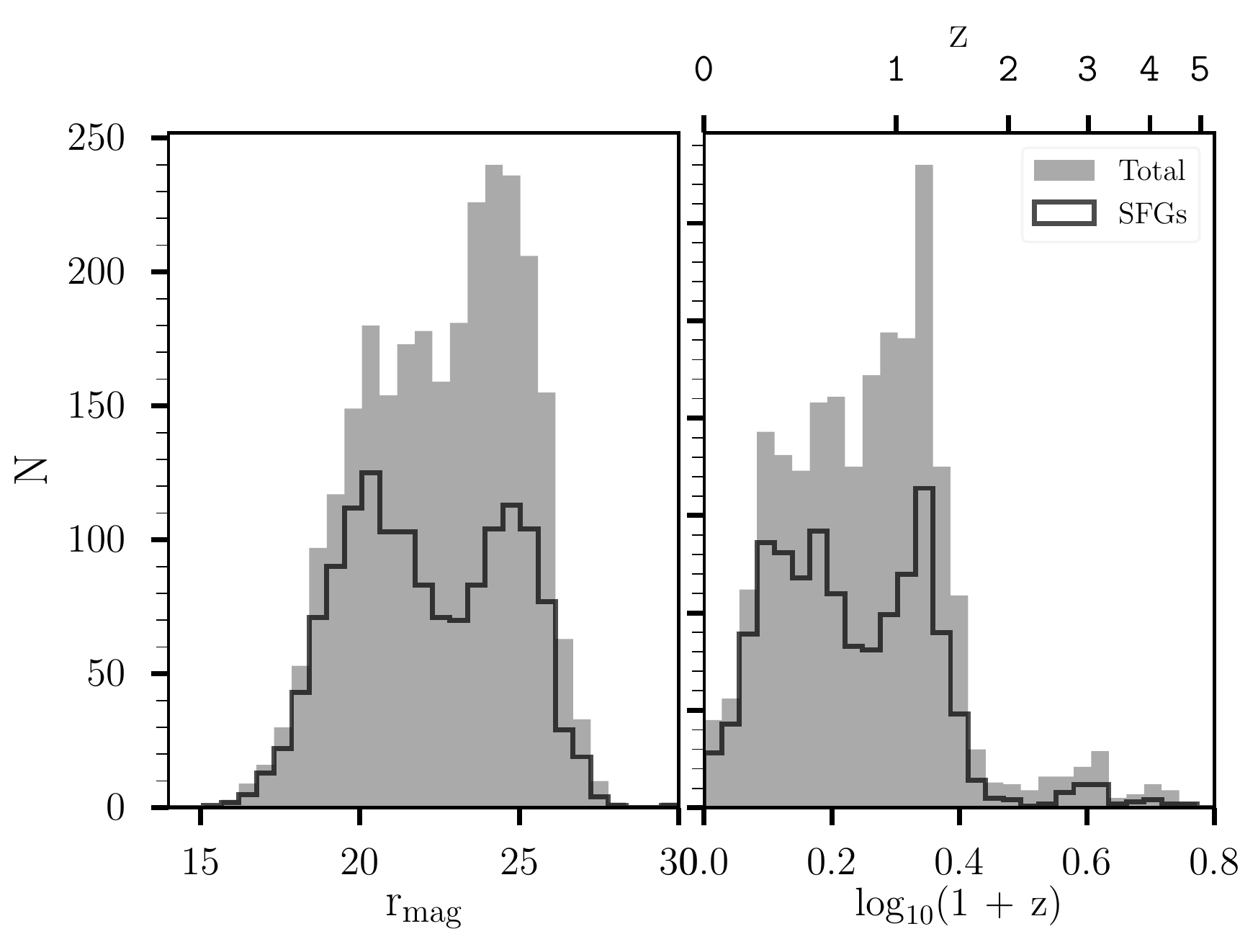}}
\caption{Distribution of the GMRT sources (grey histogram) with $\rm{r_{mag}}$ (left) and redshift (right). The distribution of SFG (black) are over-plotted.
}
\label{hist.fig} 
\end{figure}

\begin{figure}
\centering
\centerline{\includegraphics[width = 0.5\textwidth]{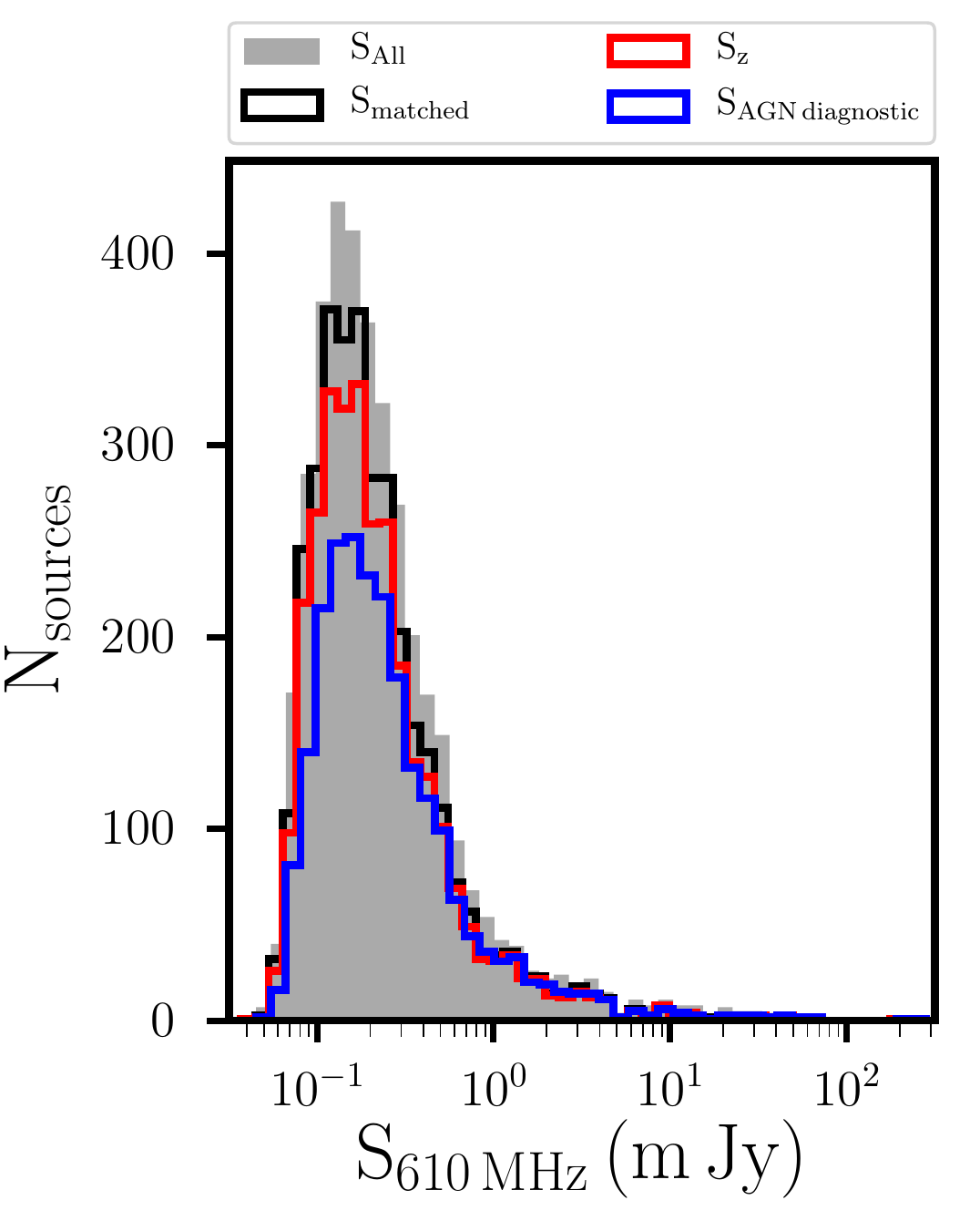}}
\caption{The distribution of 610-MHz flux densities for the entire GMRT sample of 4290 sources (light gray),radio sources that have multi-wavelength identification (black), radio sources with redshifts (red) and sources with redshifts that also have at least one diagnostic for AGN activity (blue).}
\label{comp_flux.fig} 
\end{figure}

From \citet{10.1093/mnras/stz2954}, we have a multi-wavelength match for 92\% of the sources and a redshift identification for 72\% of the sources. But we have been able to classify 39\% as SFGs and 15\% as AGNs which adds up to 54\%.  
If there is an indication of AGN activity in the source from any of the criteria adopted, then we inferred that the source is an AGN, regardless of the results from the other indicators. Thus, the sources classified as SFGs are those sources in our redshift sample that do not show evidence of AGN activity in any of the diagnostics. This may be considered as an upper limit to the population of sources in this  flux density regime whose radio emission is powered by star formation processes.  

Figure~\ref{comp_flux.fig} shows the distribution of 610-MHz flux densities.  The entire GMRT sample of 4290 sources  (i.e. light gray histogram) is represented as $\rm{S_{ALL}}$). Radio sources that have multi-wavelength identification (i.e black histogram) is represented as $\rm{S_{matched}}$, whereas radio sources with redshifts (i.e. red histogram) is represented $\rm{S_{z}}$. The sources with redshifts that also have at least one diagnostic for AGN activity (i.e. this sample is fully described in \citet{10.1093/mnras/stz2954} and Section~\ref{sfgdata.sec}, see blue histogram) is represented as $\rm{S_{AGN\,diagnostic}}$. The SFG sample used for this analysis is drawn from the latter. We use these distributions to derive the redshift success $\rm{C_{z}}$  completeness which is outlined in subsection~\ref{v_vmax.sec}.

\section{SFG Properties}\label{sfgprops.sec}
Figure~\ref{hist.fig} shows the distribution of the GMRT sources (grey histogram) with $\rm{r_{mag}}$ (left) and redshift (right). We over-plot the distribution of objects
identified as SFG in black. The redshift distribution clearly shows that 
the sample is incomplete for $\rm{r_{mag}} > 25$ and $z > 1.5$.  
This is driven by HSC/Subaru photometric redshifts, which start being incomplete at $\rm{z\,\sim\,1.3}$. \citet{2018PASJ...70S...9T} stresses that photometric redshifts should only be used at $\rm{z\,\lesssim\,1.5}$ and $\rm{i\,\lesssim\,25}$.

We note a secondary peak in our sample at $\rm{z\,\sim\,1.1}$ and $\rm{r\,\sim\,25}$.
\citet{2007MNRAS.379.1343S} identified five  candidate galaxy over densities at $\rm{z\sim1}$ across $\rm{\sim\,1\,deg^{2}}$ in the EN1 field by analysing deep field of the UK Infrared Deep Sky Survey (UKIDSS) Deep eXtragalactic Survey. 
They attributed these five overdense regions lying in a narrow redshift range as an indication of the presence of a supercluster in this field at
$z \sim 1$. Our secondary peak may be due to this supercluster.

\subsection{The redshift evolution of the IRRC}\label{IRRC}

We characterised the IR/radio correlation of our SFGs by the logarithmic ratio between the IR bolometric (8-1000 $\mu$m) luminosity and the radio luminosity $q_{\rm{IR}}$\citep{1985ApJ...298L...7H}.
 
\begin{equation}
q_{\rm IR}  = \log_{10} \left( \frac {L_{\rm IR}} {3.75 \times 10^{12}\,{\rm W} } \right ) - \log_{10} \left ( \frac{ L_{\rm radio}} {\rm W\,Hz^{-1}} \right)
\label{qIR}
\end{equation}

where $\rm{L_{IR}}$ is the total rest-frame infrared luminosity and $\rm{L_{radio}}$ is the luminosity at the radio frequency to be studied, in our case at 610 MHz and 1.4 GHz in $\rm{W/Hz}$. The 1.4 GHz luminosities were computed from the $\rm{S_{610\,MHz}}$ using a radio spectral index of -0.8.
The far-Infrared luminosities, $L_{\rm IR}$,  were derived from rest-frame integrated 8 - 1000$\rm{\mu m}$ luminosities,  estimated by \citet{2018A&A...620A..50M} using HELP photometry. They were obtained by performing SED fitting on the ultraviolet(UV)/near-infrared(NIR) to far-infrared(FIR) emission of 42,047 galaxies from the pilot HELP field: EN1. We corrected the luminosity values to our more accurate spectroscopic redshift by following the prescription presented by \cite{2017MNRAS.468.1156O}. 

We investigate the evolution of the $\rm{q_{IR}}$ parameter with redshift which is quantified by the function $\rm{q_{IR}\, \propto \, (1\, + \,z)^{\gamma}}$

\begin{figure}
\centering
\centerline{\includegraphics[width = 0.52\textwidth]{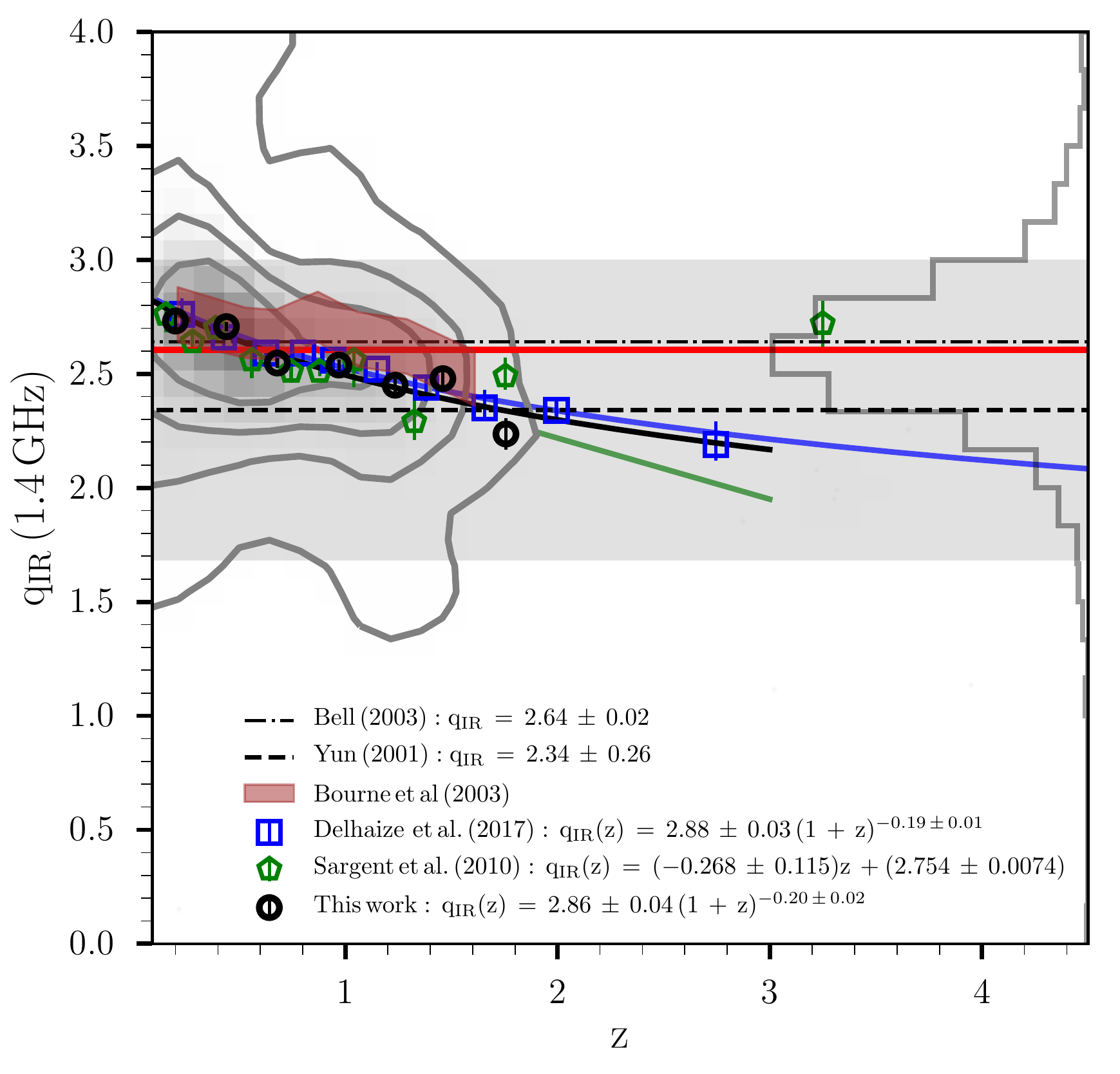}}
\caption{The $\rm{q_{IR}}$ versus redshift for SFGs. The background grey density contour represents the $\rm{q_{IR}}$ for SFGs with radio and IR detection, and redshift information. The contours levels are 1, 2, 3, and 4 $\rm{\sigma}$. The inset histogram represents the $\rm{q_{IR}}$ distribution. The median $\rm{q_{IR}}$ within each redshift bin is indicated by the open black circle. Error bars show the $\rm{1\sigma}$ dispersion calculated via bootstrap method. The solid black line shows the power-law fit to the our SFGs sample.  The horizontal shaded region represents the $\rm{\pm0.26}$ upper and lower bounds around the median value of \citet{2001ApJ...554..803Y}.
}
\label{qir.fig} 
\end{figure}

\citep{2010A&A...518L..31I,2017MNRAS.469.3468C}. We first analyse the 1.4 GHz behavior because this can be compared to the literature.
Figure~\ref{qir.fig} shows $\rm{q_{IR}}$ vs redshift. The inset histogram shows $\rm{q_{IR}}$ is scattered in a distribution with an overall median value of $\rm{q_{IR}\,=\,2.61_{-0.28}^{+0.30}}$ (see the red horizontal solid line in Figure~\ref{qir.fig}).
This $\rm{q_{IR}}$ distribution agrees well with previous literature within the errors. 
\cite{2001ApJ...554..803Y} measured a median $\rm{q_{IR}\,=\,2.34\,\pm\,0.26}$ (see the dashed horizontal line in Figure~\ref{qir.fig}), by investigating the radio counterparts to the IRAS redshift survey galaxies that are also identified in the NRAO VLA Sky Survey (NVSS) catalog. The horizontal shaded region represents the $\rm{\pm0.26}$ upper and lower bounds around the median value. \cite{2003ApJ...586..794B}  assembled a diverse sample of galaxies from the literature with far-ultraviolet (FUV), optical, infrared (IR), and radio luminosities to explore the origin of the radio-IR correlation and measured a median $\rm{q_{IR}\,=\,2.64\,\pm\,0.02}$ (see the dotted dashed horizontal line in Figure~\ref{qir.fig}).  We note that \cite{2003ApJ...586..794B} used total infrared (TIR) luminosities, but \cite{2001ApJ...554..803Y} value is based on far-infrared (FIR) luminosities . \cite{2017A&A...602A...4D} showed that this usually results in lower median values.
We split the data into  seven redshift bins, Table~\ref{qir.tab} presents the number of sources, median value of z and $\rm{q_{IR}}$ for star-forming galaxies in each redshift bin.
By fitting a power-law function to the median values of $\rm{q_{IR}}$, weighting by the uncertainty, we find a significant  variation of $\rm{q_{IR}}$ with redshift: $\rm{q_{IR}\,=\,2.86\pm0.04(1\,+\,z)^{-0.20\pm0.02}}$. The errors here are the $\rm{1\sigma}$ uncertainty from the power-law fit.
Our result is in good agreement with \cite{2017A&A...602A...4D}, who
carried out double-censored survival analysis (following \cite{2010ApJ...714L.190S}) to calculate the median $\rm{q_{IR}}$ values (and associated 95\% confidence intervals) for their samples in  redshift bins. To get $\rm{q_{IR}}$ they converted their 3 GHz luminosities to 1.4 GHz ones using a spectral index of 0.7. They reported a slightly higher but statistically significant variation of $\rm{q_{IR}}$ with redshift: $\rm{q_{IR}\,=\,2.88\pm0.03(1\,+\,z)^{-0.19\pm0.01}}$ from a highly sensitive 3 GHz observations with the Karl G. Jansky Very Large Array (VLA) and infrared data from the Herschel Space Observatory in the 2~deg$^2$ COSMOS field.  Despite the fact that we do not follow the survival analysis approach (see \citealt{1993A&A...277..114S,2017A&A...602A...5N,2018A&A...620A.192C,2018MNRAS.475..827M}), we nevertheless get results in good agreement, implying that our analysis is not significantly biased.

\cite{2015A&A...573A..45M} found a moderate but statistically significant redshift evolution $\rm{q_{IR}(z)\,=\,2.35\,\pm\,0.08(1\,+\,z)^{-0.12\,\pm\,0.04}}$ using deep FIR luminosities from the \textit{Herschel} Space Observatory \citep{2010A&A...518L...1P} and deep radio 1.4 GHz VLA observations.
\cite{2017MNRAS.469.3468C} measured the redshift evolution of the infrared-radio correlation (IRC) for SFG sample obtained with Low Frequency Array (LOFAR) at 150 MHz and found that the ratio of total infrared to 1.4 GHz data  of the Bootes field decreases with increasing redshift given by: $\rm{q_{IR}\,=\,2.45\pm0.04(1\,+\,z)^{-0.15\pm0.03}}$. 

Under our assumption of a fixed spectral index $\alpha = -0.8$, $\rm{q_{IR}\,(610\,MHz)}$ is given by the simple conversion $\rm{q_{610\,MHz}\,=\,q_{1.4GHz}\,-\,0.29}$. We thus report a median $\rm{q_{IR}\,(610\,MHz)\,=\,2.32}$, and $\rm{q_{610\,MHz}\,=\,2.57\pm0.04(1\,+\,z)^{-0.120\pm0.02}}$ as a function of redshift.


\begin{table}
 \centering
 \caption{The number of sources, median value of z and $\rm{q_{IR}}$ for star-forming galaxies in each redshift bin.}
 \begin{tabular}{|c|c|c|}
 \hline
 \hline
$\rm{z}$    & median($\rm{z}$) & $\rm{q_{IR}(1.4\,GHz)}$\\

 \hline
 $\rm{0.002\,-\,0.282}$&0.20   &$\rm{2.73\,\pm\,0.03}$\\    
 
 $\rm{0.282\,-\,0.562}$&0.44    &$\rm{2.71\,\pm\,0.02}$\\    
 
 $\rm{0.562\,-\,0.842}$&0.68     &$\rm{2.57\,\pm\,0.02}$\\    
 
 $\rm{0.842\,-\,1.122}$&0.97    & $\rm{2.54\,\pm\,0.02}$ \\   
 
 $\rm{1.122\,-\,1.402}$&1.23    & $\rm{2.45\,\pm\,0.03}$ \\   
 
 $\rm{1.402\,-\,1.682}$&1.46    & $\rm{2.48\,\pm\,0.04}$ \\   
 
 $\rm{1.682\,-\,1.962}$&1.80    & $\rm{2.24\,\pm\,0.07}$\\    
 
 
 

\hline
\end{tabular}
\label{qir.tab} 
\end{table} 

\section{Radio luminosity function }\label{rlf.sec}

\subsection{Sample selection}\label{sample.sec}

\begin{figure}
\centering
\centerline{\includegraphics[width = 0.52\textwidth]{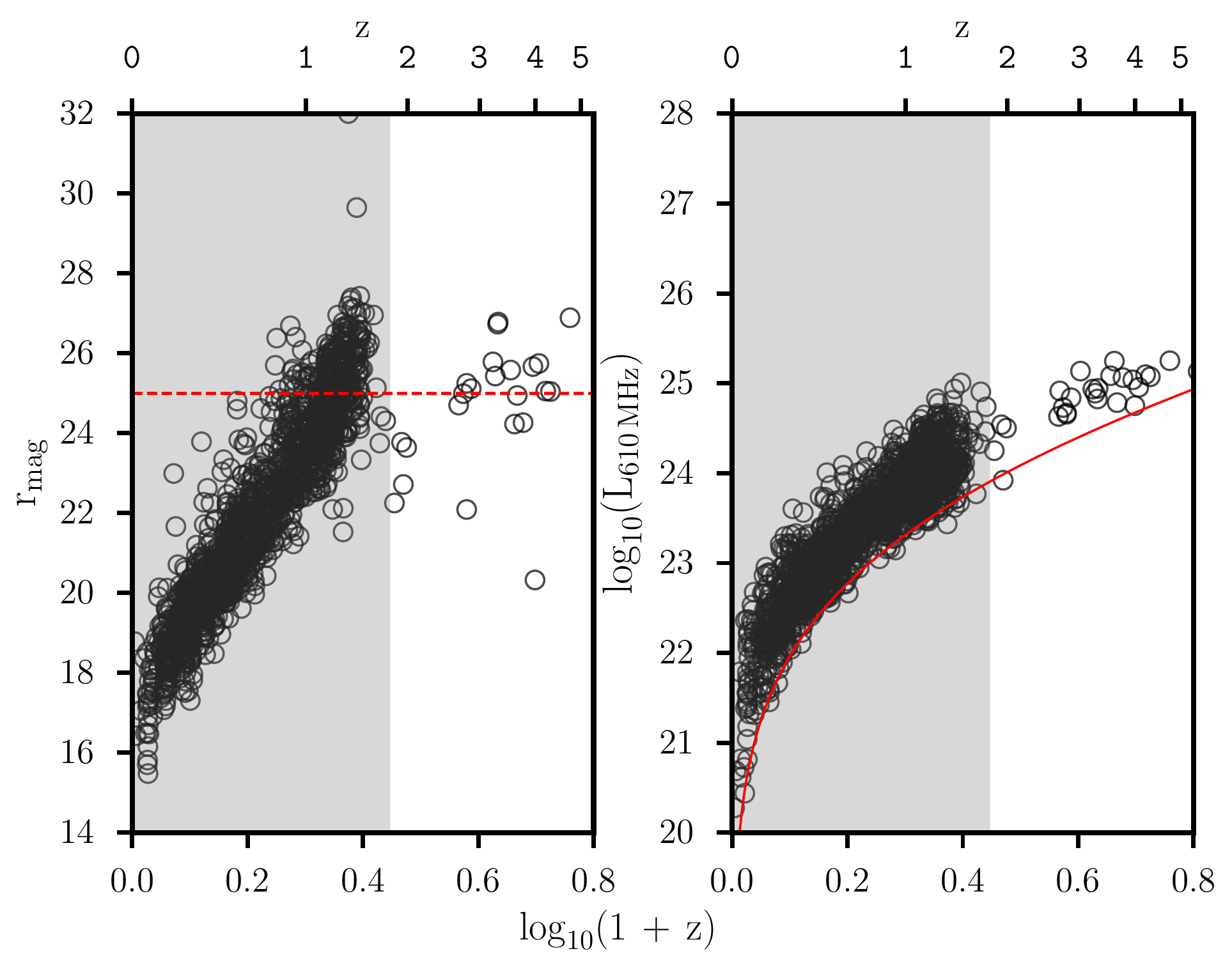}}
\caption{
$\rm{r_{mag}}$ versus redshift for the SFG sample with redshift estimates and $\rm{r_{mag}}$ limit of 25 (the dashed horizontal red line) (left panel). 610 MHz luminosity versus redshift for the GMRT SFG (open black circles) sample with redshift and  $\rm{r_{mag}}$ limit of 25 (right panel). The luminosity limit implied by the GMRT sensitivity is shown by the solid red curve.  
}
\label{rmag.fig} 
\end{figure}

To study the evolution of the Radio Luminosity Function (RLF) we limit our sample to SFGs with $\rm{r_{mag}\,lim\,=25}$ and $\rm{0.002<\,z\,<\,1.5}$, making 1291 SFGs in total.  We choose $\rm{r_{mag}\,lim\,=25}$ to maximise the number of sources that we can use to calculate the luminosity function.  Table~\ref{sel_crit.tab} presents a summary of the number and percentage of all the SFGs with spectroscopic and photometric redshifts (a). The number and percentage of the SFGs that satisfies the selection for computing the luminosity function (b). 
The $\rm{r_{mag}}$ versus redshift for the SFG sample with redshift estimates and $\rm{r_{mag}}$ limit of 25 (see the dashed horizontal red line) is plotted in the left panel of Figure~\ref{rmag.fig}. The right panel of this plot shows the 610 MHz luminosity versus redshift for the GMRT sample with redshift and limiting magnitude of r = 25.

\begin{table}
\begin{subtable}{0.57\textwidth}

\centering
\caption{All SFGs (1685 sources).}
\begin{tabular}{|c|c|c|}

\toprule
 & Number & Percentage \\ \midrule
$\rm{z_{phot}}$               & 1189                      & 70.5                  \\
$\rm{z_{spec}}$              & 496                      & 29.5                  \\
\bottomrule

\end{tabular}

\end{subtable}

\begin{subtable}{0.57\textwidth}
\centering
\caption{SFGs and selection criteria  (1291 sources, \\ 
i.e  $\rm{SFGs\,\wedge\,r\,=\,25\,\wedge\,0.002\,<\,z<\,1.5}$).} 
\begin{tabular}{|c|c|c|}
\toprule
 & Number & Percentage \\ \midrule
$\rm{z_{phot}}$               & 834                      & 64.6                  \\
$\rm{z_{spec}}$             & 457                      & 35.4                  \\

\bottomrule
\end{tabular}
\end{subtable}
\begin{tablenotes}
 \item $\rm{z_{phot}}$ - photometric redshift.
\item  $\rm{z_{spec}}$ - spectroscopic redshift.
 \end{tablenotes}
\caption{Summary of the number and percentage of all the SFGs with spectroscopic and photometric redshifts (a). The number and percentage of the SFGs that satisfies the selection for computing the luminosity function (b).}
\label{sel_crit.tab} 
 
\end{table}

\subsection{\texorpdfstring{$\rm{V/V_{max}}$}{Lg} Statistic}\label{v_vmax.sec} 
In order to assess the evolution in the comoving space density of radio sources we use the non-parametric $\rm{V/V_{max}}$ method (\citealt{1968MNRAS.138..445R,1968ApJ...151..393S}).
$\rm{V_{max}}$ is the volume over which the galaxy could have been observed given the selection limits. 
It allows the incorporation of additional selection criteria. 

For a uniform distribution, the value of $\rm{V/V_{max}}$ will be uniformly distributed between 0 and 1. Thus for such a sample the mean value is $\rm{\langle(V/V_{max})\rangle\, = \, 0.5\,\pm\,(12N)^{-1/2}}$, where N is the number of objects in the sample. $\rm{\langle(V/V_{max})\rangle\, > \, 0.5}$ indicates that the sources are biased towards larger distances, or an increase of the space density with redshift. $\rm{\langle(V/V_{max})\rangle\, < \, 0.5}$ indicates a deficiency in high redshift sources, or a decline in the space density with redshift. A constant comoving population is given by 
$\rm{\langle(V/V_{max})\rangle\, =\, 0.5}$ (\citealt{doi:10.1111/j.1365-2966.2004.07981.x,2008A&A...490..879T,doi:10.1111/j.1365-2966.2010.18191.x,2013MNRAS.436.1084M,2016MNRAS.457..730P}).

Our sample is a matched radio/optical sample, thus we take into account both the optical
and radio limits of the surveys, where $\rm{V_{max}}$, the final maximum observable
volume, is taken as the minimum from the optical and radio $\rm{V_{max}}$ for each source:

\begin{equation}
\rm{V_{max}\, = \,  min(V_{max,radio}, V_{max,optical})}
\end{equation}

Where $\rm{V_{max,radio}}$ and $\rm{V_{max,optical}}$ represent the maximum observable volumes
of the source in the radio and optical surveys respectively and are shown below:
\begin{equation}\label{v_radio}
\rm{V_{max,radio} = \sum\limits_{i=1}^n V_{max,radio,i}(z_{max,radio,i})\times\,\mathcal{C}_{i}}
\end{equation}

\begin{equation}\label{v_optical}
\rm{V_{max,optical} = \sum\limits_{i=1}^n V_{max,optical,i}(z_{max,optical,i})\times\,\mathcal{C}_{i}}
\end{equation}

$\rm{V_{max,radio}}$ and $\rm{V_{max,optical}}$ were  computed from $\rm{z_{max,radio}}$ and $\rm{z_{max,optical}}$ as shown in equations~\ref{v_radio} and \ref{v_optical} above.  This is in a single redshift bin and that the sum goes over all galaxies in a given redshift bin. The
$\rm{z_{max,radio}}$ and $\rm{z_{max,optical}}$ represent the maximum observable redshifts of the source in the radio and optical surveys 
respectively. The k-correction to the $\rm{V_{max,radio}}$ is a power law.
We estimate $\rm{z_{max,optical}}$ by running \texttt{ kcorrect} \citep{2003AJ....125.2348B} which redshifts the best fitting SED template from the photometric redshift estimation procedure and determine the redshift where the template
becomes fainter than our imposed a limiting magnitude of r = 25. The derivation of the radio completeness $\rm{C_{f}}$  is given by $\rm{\epsilon(s)}$ (see \citet{10.1093/mnras/stz2954}). The $\rm{\epsilon(s)}$ is the probability that a source with true flux density,  $\rm{s}$, will result in a detection. We measured this by inserting 3000 artificial point sources at a given true flux density at random positions into the residual map with the original sources removed. These sources populate the image with the same background noise and rms properties as the original source finding. We stressed in \citet{10.1093/mnras/stz2954} that the field of view effect dominates the curve in the radio completeness correction (see Figure 7, \cite{10.1093/mnras/stz2954}) since the analysis is incorporating the varying sensitivity limit across the field of view due to the GMRT primary beam. This represents the completeness of the radio source catalogue versus true flux density (see \citealt{10.1093/mnras/stz2954}). In order to correct for the redshift incompleteness, we divided the distribution for the entire sample (i.e. light gray histogram see Figure~\ref{comp_flux.fig}) by the distribution for the sources with at least one diagnostic for AGN activity (i.e. blue histogram). We then corrected the RLF with this redshift success completeness, $\rm{C_{z}}$.  This redshift incompleteness in our sample is mostly due to sources not being detected in the optical wavelength range, so that we cannot compute a reliable (photometric) redshift.
We define the completeness correction factor $\rm{\mathcal{C}_{i}}$ (i.e see equations~\ref{v_radio} and \ref{v_optical}) as:
\begin{equation}\label{completeness}
\rm{\mathcal{C} = C_{z}\,\times\,C_{f}}
\end{equation}

$\rm{C_{z}}$ is the redshift success completeness and $\rm{C_{f}}$ is the completeness of the radio catalog.

\begin{figure}
\centering
\centerline{\includegraphics[width = 0.5\textwidth]{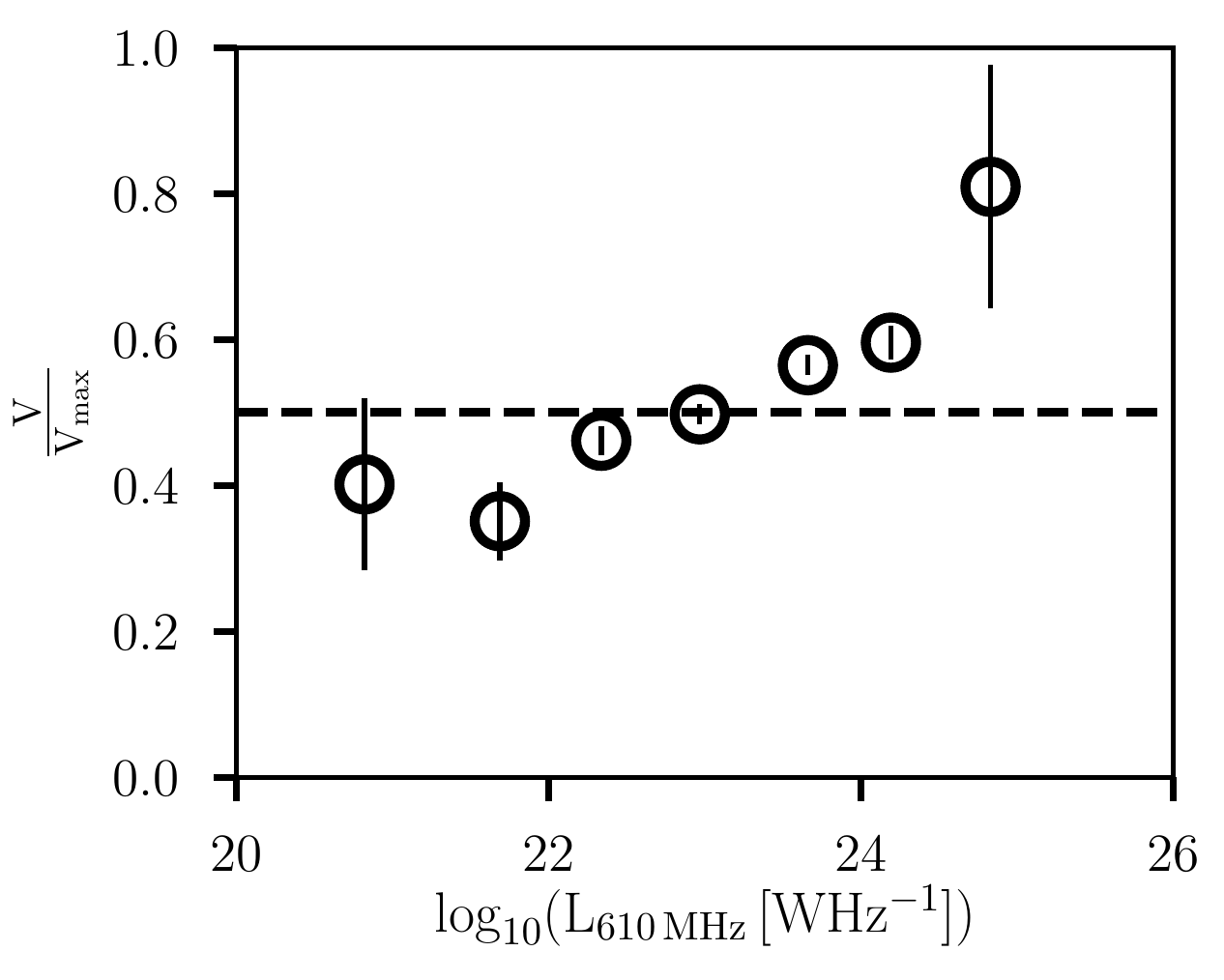}}
\caption{The $\rm{V/V_{max}}$ statistic as a function the radio luminosity for SFG
out to $\rm{z\, =\, 1.5}$ for the GMRT data (open black circles). 
}
\label{vover_vmax.fig} 
\end{figure}

In Figure~\ref{vover_vmax.fig} and Table~\ref{vover_vmax.tab} we show the mean $\rm{V/V_{max}}$ statistic in bins of radio luminosity in the range $\rm{10^{20}\, <\, L_{610MHz}\, < \,10^{25}\,WHz^{-1}}$ for our SFG sample (see open black circles). 
For each bin we provide the number of sources ($\rm{N}$) in the bin, mean $\rm{\langle V/V_{max} \rangle}$, mean redshift ($\rm{\langle\,z\,\rangle}$). The $\sigma\,=\,1/\sqrt{12N}$ \cite{1980ApJ...235..694A} are the statistical errors derived from the sample size. The dashed horizontal line shows the median $\rm{\langle V/V_{max} \rangle}$. We calculate $\rm{\langle V/V_{max} \rangle}$ to be $\rm{0.51\,\pm\, 0.06}$ for our SFGs.
 This value is not  significantly different from 0.5, given the error of 0.06  thus averaged over luminosity there is no overall evidence for evolution in 
the number density of SFGs.
However there is a clear trend of $\rm{V/V_{max}}$ with radio luminosity.
At faint radio luminosities $\rm{L_{610MHz}\, < \,10^{23}\,WHz^{-1}}$ the
values are below 0.5 indicating a higher space density at low redshift. 
Conversely for high luminosity, $\rm{L_{610MHz}\, >\, 10^{23}\,WHz^{−1}}$ there
is evidence for positive evolution or higher space density at higher 
redshift. .
The strong evolution of the high luminosity radio sources was also detected by 
\citet{doi:10.1111/j.1365-2966.2004.07981.x}). 
\citet{doi:10.1111/j.1365-2966.2010.18191.x} also found evidence of this strong evolution at high luminosities although their result is at a lower statistical significance due to the small size of their sample.
Taken together the results indicate strong luminosity evolution with overall number density constant with redshift but more high luminosity sources and fewer low-luminosity objects at higher redshift.

\begin{table}
\caption{ The $\rm{V/V_{max}}$ statistic as in radio luminosity bins for SFGs.}
\centering
\begin{tabular}{|c|c|c|c|}
\hline
\hline
Median Luminosity&Number&$\rm{V/V_{max}}$&$1/\sqrt{12N}$  \\
 
 $\mathrm{\log_{10}(L_{610\,MHz}\,[WHz^{-1}])}$ &(SFG)&(SFG)&(SFG)\\
\hline
20.82   & 6&0.446&0.12       \\

21.72   &29&0.258&0.05      \\

22.34    &211&0.456&0.02      \\

22.97    & 433&0.494&0.01       \\

23.66    & 442&0.569&0.01       \\

24.33    & 162&0.583&0.02      \\

24.93  & 3&0.729&0.17       \\
\hline
\end{tabular}
\label{vover_vmax.tab} 
\end{table} 

\subsection{Derivation of the Radio luminosity function (RLF)}
We derive the radio $\rm{LF \, (\Phi)}$  for our GMRT sample in five redshift bins using the standard $\rm{\frac{1}{V_{max}}}$ method
 \citep{1968ApJ...151..393S}.

As the GMRT mosaics have non-uniform sensitivity, the effective area of the survey changes as a function
of the flux limit. The volume of space available to a source of a given luminosity $\rm{V_{max, radio}}$ (L)
has to be calculated by taking into account the variation of survey area as a function of flux density limit.
The RLF for a given luminosity bin is given by:

 \begin{equation}\label{phi.eqn}
 \centering
\rm{ \Phi_{z} (L)\, = \, \sum\limits_{i=1}^n \frac{1}{V_{max,i}}  \, \pm \sqrt { \sum\limits_{i=1}^n \frac{1}{{V_{max,i}}^{2} } }}
\end{equation}

where $\rm{\Phi_{z}(L)}$ is the density of sources in $\rm{Mpc^{-3}dex^{-1}}$.

\section{Cosmic evolution of the SFG radio luminosity function}\label{evol.sec}

In this section we explain the reasoning behind adopting the analytic form of our local luminosity function at 610 MHz to fit our data. We further describe how the evolution of SFG luminosity function out to $\rm{z\sim\,1.5}$ is constrained.

\subsection{The local RLF}\label{loc_evo.sec}

\begin{figure}
\centering\centerline{\includegraphics[width = 0.5\textwidth]{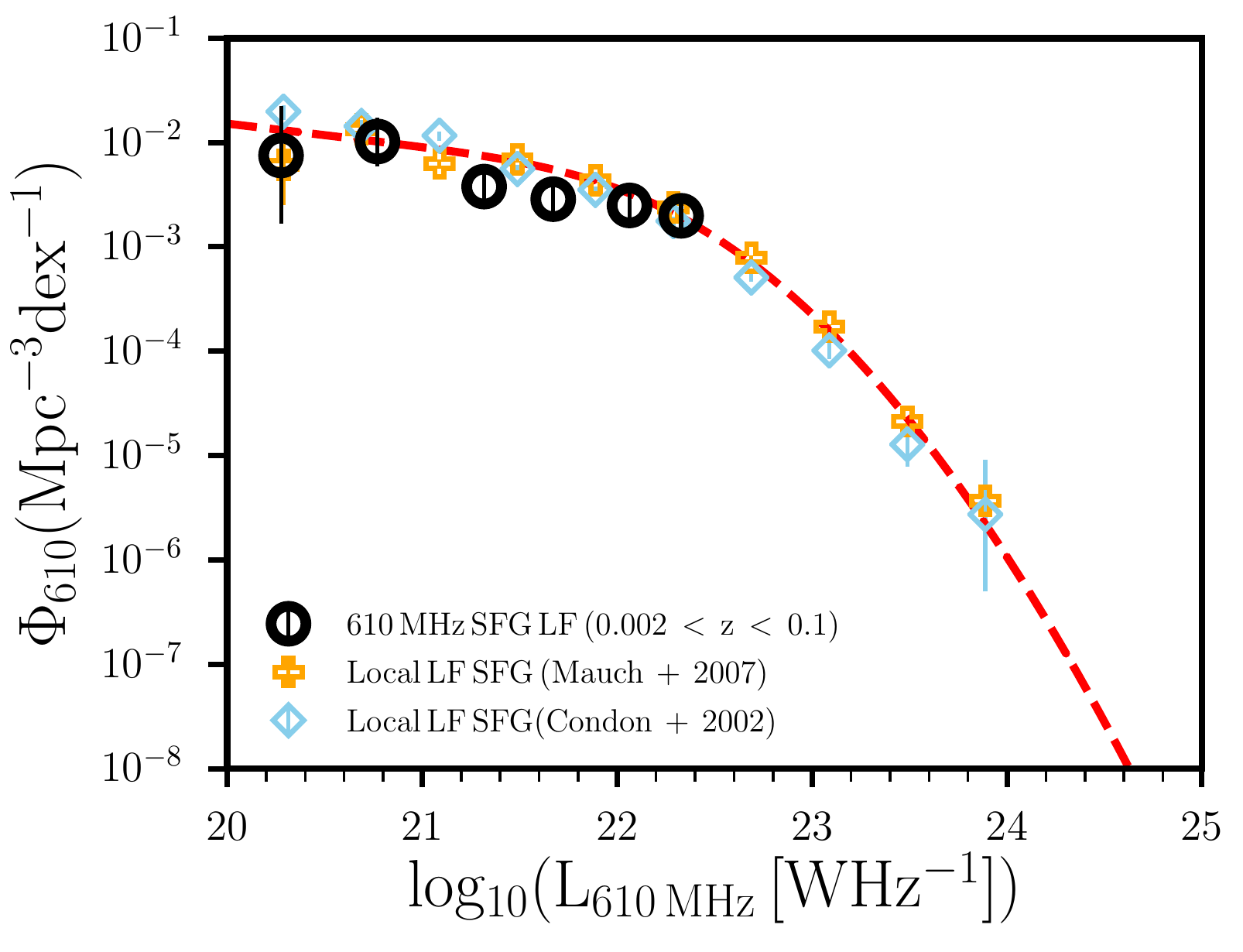}}
\caption{The local 610 MHz SFG luminosity function.
The yellow plus and blue stars represents \citet{2007MNRAS.375..931M} and \citet{2002AJ....124..675C} SFG volume densities scaled to 610 MHz using an $\mathrm{\alpha\,=\,-0.8}$. The dashed red line is the combined
analytic fit performed by \citet{2017A&A...602A...5N} to the local radio LF of SFGs from several surveys with
different observed areas and sensitivities. }
\label{locLF.fig} 
\end{figure}

Figure~\ref{locLF.fig} presents the local 610 MHz SFG luminosity function shown as open black circles. The sample is truncated
at $\rm{z\,<\,0.1}$ to minimize the effects of evolution.
The yellow plus and blue stars represents \citet{2007MNRAS.375..931M} and \citet{2002AJ....124..675C} SFG volume densities scaled to 610 MHz using an $\mathrm{\alpha\,=\,-0.8}$. The dashed red line is the
analytic fit to the local 610 MHz SFG data. We also show the 610 MHz RLF for SFGs in the redshift range $\rm{0.002\, < \,z \,<\,1.5}$ shown in open black circles in Figure~\ref{lumfunc_all.fig}.

An analytic function of the type described by \citet{1990MNRAS.242..318S}

\begin{equation}\label{eq:lognorm}
\rm{\Phi_{o}(L) = \Phi_{\star}\,\Bigg(\frac{L}{L_{\star}}\Bigg)^{1-\alpha}exp \Bigg[\frac{-1}{2\sigma^{2}}\log^{2} \Bigg(1+\frac{L}{L_{\star}}\Bigg)\Bigg] }
\end{equation}

where the $\rm{L_{\star}}$ parameter describes the position of the turnover of the power-law plus log-normal distribution, $\rm{\Phi_{\star}}$ is  the normalization, $\rm{\alpha}$ and $\rm{\sigma}$ are the faint and bright ends of the distribution, respectively. 

To obtain the analytic form of the local luminosity function that is used throughout this work we use the best fit parameters from  \citet{2017A&A...602A...5N}, who combined data from  both wide and deep surveys  to properly constrain both the faint and the bright end of the local LF form \citet{2002AJ....124..675C}, \citet{2005MNRAS.362...25B}, \citet{2007MNRAS.375..931M} data using the form given in equation~\ref{eq:lognorm}. 
The best fit parameters obtained by \citet{2017A&A...602A...5N}, which we use throughout this work are
$\rm{\Phi_{\star}\,=\,3.55\times10^{3} Mpc^{-3}dex^{-1}}$, $\rm{L_{\star}\,=\,1.85 \times 10^{21} WHz^{-1}}$, $\rm{\alpha\,=\,1.22}$, $\rm{\sigma\,=\,0.63}$. 

\subsection{RLF as a function of z}
\begin{figure*}
\centering
\centerline{\includegraphics[width =  0.9\textwidth]{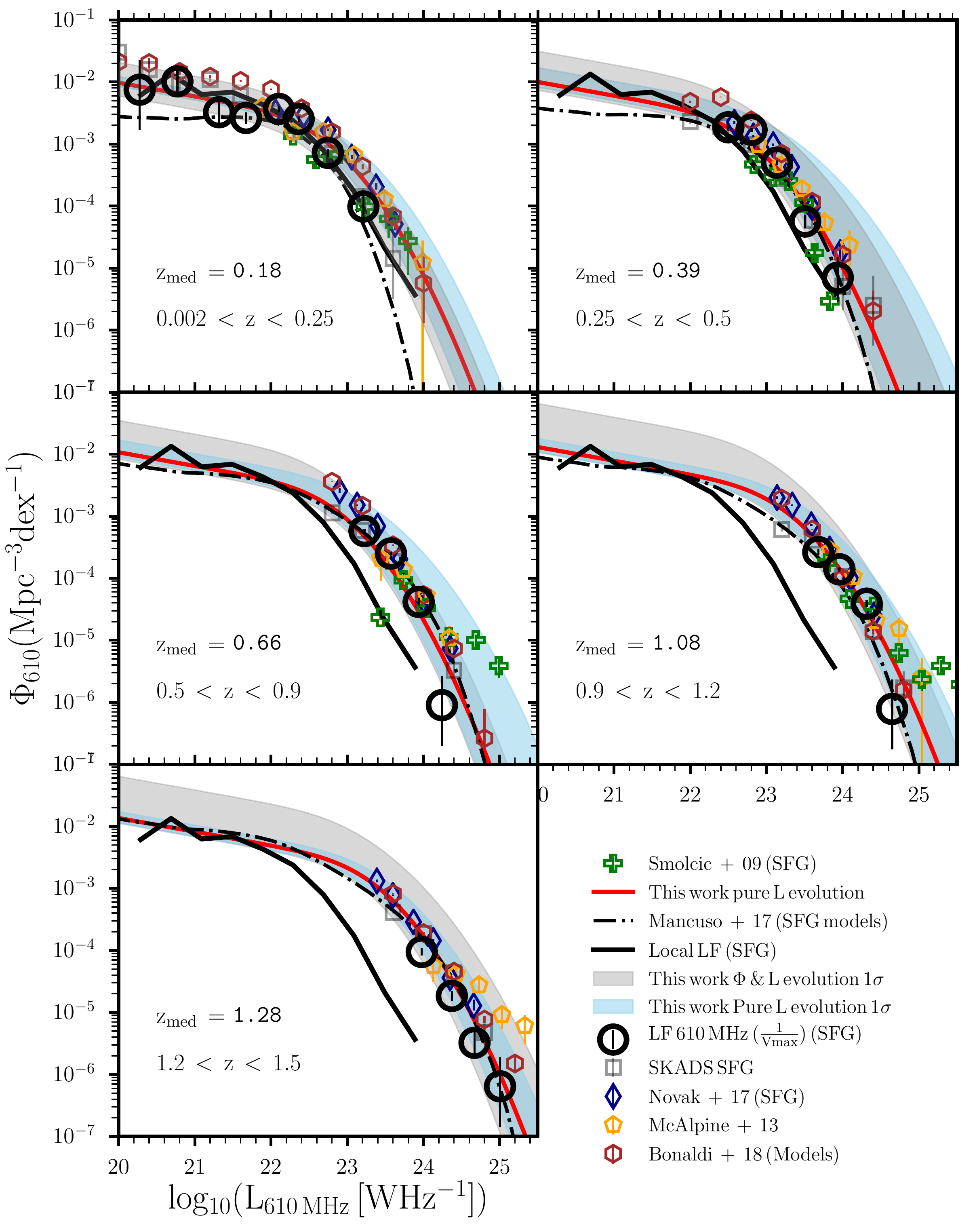}}
\caption{Radio luminosity functions of SFGs at $\rm{\nu}\, = \,610\,MHz$ in different redshift bins (black open circles). The black dashed lines in each panel are the SFG models from \citet{Mancuso2017}. The black squares represent the total SFG LF from the semi-empirical simulation of the SKA \citep{2008MNRAS.388.1335W}. Luminosity functions computed for SFGs from the T-RECS \citep{Bonaldi2019} simulations are shown as open brown diamonds. The local radio luminosity function of \citet{2007MNRAS.375..931M} is shown for reference as a solid black line in each panel. Scaled down luminosity functions from 1.4 GHz to 610 MHz by \citet{2009ApJ...690..610S}, \citet{2013MNRAS.436.1084M} and \citet{2017A&A...602A...5N} are shown as green pluses, orange pentagons and blue diamonds respectively in each panel. The solid red lines in each panel corresponds to the median values of the MCMC samples and the shaded regions correspond to the 68\% confidence region of the PLE fit (skyblue) and also the combination of PDE and PLE fitting (grey) to the samples. The redshift range  and the median redshift are shown in each panel. Error bars are determined using the prescription of \citet{1986ApJ...303..336G}.}
\label{lf_610_sfg} 
\end{figure*}

We compare our results with literature values of SFG LF derived at 1.4 GHz and scaled down to 610 MHz assuming $\mathrm{ \alpha\,=\,-0.8}$ to check the robustness of our LF. Figure~\ref{lf_610_sfg} presents the radio luminosity functions of SFGs at $\rm{\nu}\, = \,610\,MHz$ in different redshift bins (black open circles). 
Scaled down luminosity functions from 1.4 GHz to 610 MHz by \citet{2009ApJ...690..610S}, \citet{2013MNRAS.436.1084M} and \citet{2017A&A...602A...5N} are shown as green pluses, orange pentagons and blue diamonds respectively in each panel. 

We compare with  LFs derived from \citet{2008MNRAS.388.1335W} semi-empirical simulation of the SKA and \citet{Mancuso2017} models. The \citet{Mancuso2017} models were obtained by following the model-independent approach by \citet{Mancuso2016a,Mancuso2016b}.
These models are based on two
main ingredients: (i) the redshift-dependent SFR functions inferred from the latest UV/far-IR data from HST/Herschel and related statistics of strong gravitationally lensed sources,
and (ii) deterministic tracks for the co-evolution of star formation and BH accretion in an individual galaxy, determined from a wealth of multiwavelength observations (see \cite{Mancuso2017}).
 We also compare to SFG models (see open brown diamonds in Figure~\ref{lf_610_sfg}) from the Tiered Radio Extragalactic Continuum Simulation (T-RECS) by \cite{Bonaldi2019} who modeled the corresponding sub-populations, over the 150 MHz - 20 GHz range. 
Our results concurs with the results of these models from literature, especially to the \citet{Mancuso2017} and \cite{Bonaldi2019} models at high luminosities.
 Note, however, that in the  first two redshift bins the faint end of the \citet{Mancuso2017} models is lower than that of our SFG LF. Also, the faint end of the \cite{Bonaldi2019} and \citet{2008MNRAS.388.1335W} models are higher than our measured SFG LF.
 
The breakdown of the luminosity, number density ($\rm{\Phi_{610}(Mpc^{-3}dex^{-1})}$) and the number of sources in each redshift bin is presented in Table~\ref{LF_bins.tab}. Our data have small Poissonian error bars due to the relatively large number of sources in each bin and as such the errors do not reflect all possible systematic effects.

\subsection{RLF Evolution}

Following \citet{2017A&A...602A...5N}, we assume that the shape of the LF remains unchanged at all observed cosmic times and  only allow the position of the turnover and the normalization to change with redshift.  We used the Markov chain Monte Carlo (MCMC) algorithm  module \textsc{emcee} \citep{2013PASP..125..306F}, implemented in the \textsc{lmfit} Python package \citep{https://doi.org/10.5281/zenodo.11813} to perform a multi-variate fit to the data. We fit all redshift slices for evolution assuming two scenarios for the LF, one in which the luminosity of the radio sources is fixed and undergoes pure density evolution parametrized (PDE) by

\begin{equation}\label{eq:puredens} 
\rm{\Phi_{z}(L) = (1+z)^{k_{D}}\Phi_{o}(L)}
\end{equation}

and another in which the number density of radio sources is fixed and the population undergoes pure luminosity evolution:

\begin{equation}\label{eq:purelum}
\rm{\Phi_{z}(L) = \Phi_{o}\Bigg(\frac{L}{(1+z)^{k_{L}}}\Bigg)}
\end{equation}

where $\rm{\Phi_{z}(L)}$ is the LF at redshift z, $\rm{\Phi_{o}(L)}$ the normalization of the local LF, and $\rm{k_{D}}$ and $\rm{k_{L}}$  represent pure density and pure luminosity evolution parameters, respectively and denotes the strength of the evolution.
Both the PLE and PDE models are common in the literature (e.g. see \citealt{1984ApJ...287..461C,2002MNRAS.329..227S,2007MNRAS.381..211S,2013MNRAS.432...23G,2013MNRAS.436.1084M}). Studies have shown that true evolution might be a combination of both of these extremes (see, e.g. \citealt{2016MNRAS.460.3669Y,2016ApJ...820...65Y,2017A&A...602A...5N}). The best fit evolution parameters
for each redshift bin obtained with this procedure are presented
in Table~\ref{evo.tab}. 
The \textsc{lmfit} Python package first does the fitting by performing a non linear least-squares  $\rm{\chi^{2}}$
minimization to obtain the best fit $\rm{k_{L}}$ and $\rm{k_{D}}$ parameters. The \textsc{emcee} is then implemented to calculate the  probability distribution for the parameters. From this we get the medians of the probability distributions and a 1$\rm{\sigma}$ quantile, estimated as half the difference
between the 15.8 and 84.2 percentiles. 

We also fit a continuous model the redshift dependence
of the evolution parameters by adding a redshift dependent term to the $\rm{k_{L}}$, and $\rm{k_{D}}$ parameters in Equation~\ref{eq:puredens} and Equation~\ref{eq:purelum} (e.g. see \citealt{2017A&A...602A...5N,2017A&A...602A...6S,2018A&A...614A..47N,2018A&A...620A.192C}). We fit a simple linear redshift dependent evolution model to all SFG luminosity functions in all redshift bins simultaneously given by:
\begin{equation}\label{eq:cont_model} 
\rm{\Phi(L,z)\,=\,(1+z)^{(k_{D}+z\beta_{D})}\times\,\Phi_{0}\Bigg[\frac{L}{(1+z)^{(k_{L}+z\beta_{L})}}\Bigg]}.
\end{equation}
where $\rm{k_{D}}$, $\rm{k_{L}}$, $\rm{\beta_{D}}$ and $\rm{\beta_{L}}$ are the various evolution parameters.  Equation~\ref{eq:cont_model} considers the case with both density and luminosity evolution combined plus redshift dependence (i.e four free parameters ).
We test pure density and pure luminosity evolution together via the  procedure described above.
The 68\% confidence region by combining PDE and PLE fitting to the samples are shown with grey shaded are Figure~\ref{lf_610_sfg}.

\begin{figure}
\centering
\centerline{\includegraphics[width = 0.55\textwidth]{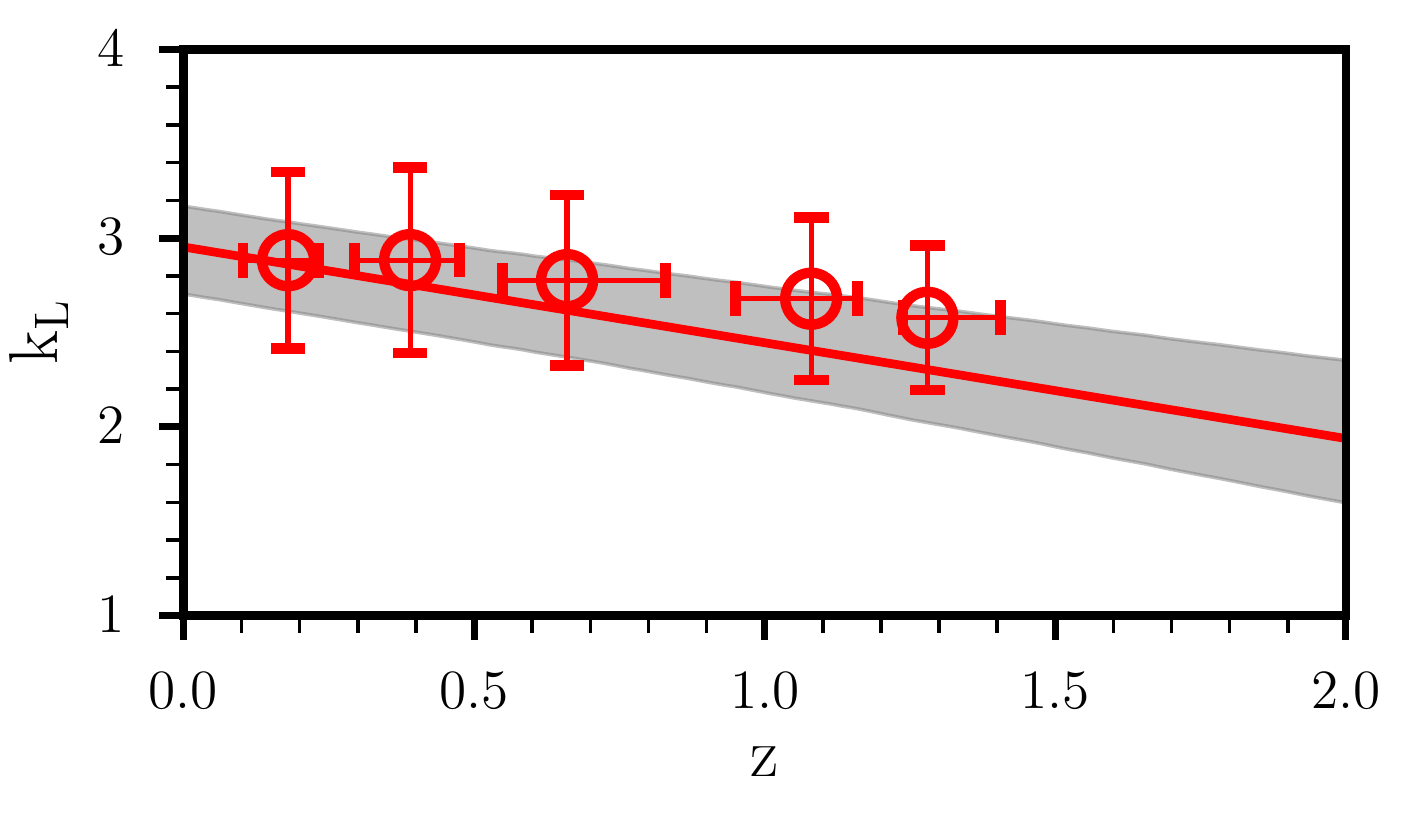}}
\caption{Parameters obtained from fitting PLE model to the SFG luminosity functions. Open red circles show the evolution parameters obtained
from fitting the assumed analytic form of the luminosity function
in five redshift bins assuming pure luminosity 
evolution scenario (see text for details).  The vertical error bars represent the median absolute deviation (MAD) of the MCMC samples. The horizontal error bars denote the inter-quartile range (IQR) of reshift in each bin. The same
color line shows the results from the continuous fit assuming that the PLE parameter evolves linearly with redshift.}
\label{evo_params.fig} 
\end{figure}

Figure~\ref{evo_params.fig} presents best fit parameters obtained from fitting PLE model to the SFG luminosity functions. The second column of Table~\ref{evo.tab} presents values of the parameters (i.e. $\rm{k_L}$) obtained from fitting PLE model to the SFG luminosity functions
Open red circles show the evolution parameters obtained from  independently fitting the assumed analytic form of the luminosity function
in five redshift bins assuming pure luminosity 
evolution scenario whereas the same
color line shows the results from the continuous fit (i.e.  jointly fitting the RLF in the 5 redshift bins) assuming that the PLE  parameter evolves linearly with redshift by using Equation~\ref{eq:cont_model}. The vertical error bars represent the median absolute deviation (MAD) \citep{doi:10.1080/01621459.1993.10476408} of the MCMC samples. We derive $\rm{L_{610\,MHz}\,\propto\,(\,1+\,z)^{(2.95\pm0.19)-(0.50\pm0.15)z}}$ (i.e. $\rm{k_{L}\,=\,2.95\,\pm\,0.19}$) for $\rm{0.002\,<\,z\,<\,1.5}$.
With no uncertainties associated with their estimated value, \citet{2000ApJ...544..641H} have found that a PLE with $\rm{k_{L}\,=\,2.74}$ is a good representation of the evolution of their radio-selected SF galaxies.
\citet{2009ApJ...690..610S} derived  $\rm{2.1\pm0.2}$ or $\rm{2.5\pm0.1}$  for SFG depending on the choice of the local LF (i.e. \citet{2002MNRAS.329..227S} or \cite{1989ApJ...338...13C} local LFs).  \citet{2010ApJ...714.1305S}  derived $\rm{k_{L}\,=\,2.9\,\pm\,0.3 }$ by studying a sample of 1.4 GHz radio sources in the Deep SWIRE
Field (DSF), reaching a limiting flux density $\rm{\sim 13.5 \mu Jy}$ at the center of a $\rm{0.36\,deg^{2}}$ area. We can therefore compare
our results more directly with both \citet{2009ApJ...690..610S} and \citet{2010ApJ...714.1305S}, keeping in mind
that both samples reach only $\rm{z\,= \,1.3}$. The evolution we measure is slightly higher than \citet{2009ApJ...690..610S} but in good agreement with \citet{2010ApJ...714.1305S}. However, our value is significantly weaker than found in the IR band for $\rm{z\,\leq\, 1.3}$ by \citet{2009A&A...496...57M}, who modeled  the evolution of infrared luminous star-forming
galaxies as a PLE and found $\rm{k_{L}\,= \,3.6\pm0.4}$.  
\citet{2013MNRAS.436.1084M} studied the evolution of faint radio source out to $\rm{z\sim2.5}$. They found that the radio population experiences mild positive evolution out to $\rm{z\sim1.2}$ increasing their space density by a factor of $\sim3$ with SFGs driving the more rapid evolution at low redshifts, $\rm{z<1.2}$. The  \citet{2013MNRAS.436.1084M} translated to 610 MHz are shown in  yellow symbols in Figure~\ref{lf_610_sfg}
and are in good agreement with our values of SFG  luminosity functions out to $\rm{z<1.5}$. They reported $\rm{k_{L}}$ to be $\rm{2.47\pm0.12}$ for their radio-selected star-forming population which is consistent with \citet{2009ApJ...690..610S} but slightly below our measured value.
\cite{2011ApJ...740...20P} reported the radio power of SFGs evolves as $\rm{(1\, +\, z)^{2.5\,-\,2.9}}$ up to $\rm{z\,\leq\,2.3}$, their maximum redshift in their sample, which in agreement with previous determinations in the radio, IR bands and this work. Although they also reported  the evolution to be $\rm{k_{L}\,=\,3.5^{+0.4}_{−0.7}}$
for $\rm{z\,\leq\,1.3}$ or $\rm{k_{L}\,=\,3.1^{+0.8}_{−1.0}}$, when they exclude two large-scale structures in their sample.
\citet{2017A&A...602A...5N} presented a radio selected sample of star-forming galaxies
from deep VLA-COSMOS 3 GHz observations \citep{2017A&A...602A...6S} identifying 6040 galaxies, where the radio emission is not dominated by an AGN. Using
this sample they derived radio LFs up to ${z\sim5}$. The blue diamonds in Figure~\ref{lf_610_sfg} show the \citet{2017A&A...602A...5N} LFs scaled down to 610 MHz. Their results are in agreement with our luminosity functions, with the their LF constraining the high luminosity end.
By comparing their results with LFs derived using IR and UV selected samples and  checking their robustness, they  reported that their radio LF can be well described by a local LF evolved only in luminosity as
$\rm{L_{1.4 GHz}\,\propto\,(1+z)^{(3.2\pm0.2)−(0.33\pm0.08)z}}$. These previous  studies are broadly consistent with our radio derived PLE parameter and Table~\ref{LF_evo.tab} presents a summary of the comparison.

\begin{table*}
 \centering
 \caption{Comparison of the current determinations of the evolution of the radio luminosity function.}
 \begin{tabular}{|c|c|c|c|c|}
 \hline
 \hline
Reference & Field    & Wavelength &Redshift & Evolution Parameter (PLE) \\

\hline
 \citet{2004ApJ...615..209H}& - &1.4 GHz  & $\sim2.0$ & $\rm{2.7\pm0.6}$     \\
\citet{2009ApJ...690..610S}&COSMOS&1.4 GHz  & $\sim1.3$ & $\rm{2.1\pm0.2}$ OR $\rm{2.5\pm0.1}$    \\
\citet{2010ApJ...714.1305S}& DSF &1.4 GHz  & $\sim1.3$ & $\rm{2.9\pm0.3 } $ \\
\cite{2011ApJ...740...20P}&CDFS&1.4 GHz& $\leq1.3$& $\rm{3.5^{0.4}_{-0.7}}$\\
\cite{2011ApJ...740...20P}&CDFS&1.4 GHz& $\sim2.3$& $\rm{2.89^{+0.10}_{-0.15}}$\\
 \text{\citet{2017A&A...602A...5N}}& COSMOS &1.4 GHz & $\rm{\sim 5.0}$ &$\rm{(3.2\pm0.2)\,-\,(0.33\pm0.08)z}$\\
This work &EN1&610 MHz  & $\sim1.5$ & $\rm{(2.95\pm0.19)\,-\,(0.50\pm0.14)z}$ \\

 \hline   
\end{tabular}
\begin{tablenotes}
\item [a] COSMOS - Cosmological Evolution Survey. 
\item [b] CDFS - Chandra Deep Field South.
\item [c] DSF - Deep SWIRE Field.
\end{tablenotes}
\label{LF_evo.tab} 
\end{table*}

\section{The Cosmic Star Formation History traced by the low-frequency SFG population}\label{csfh.sec}
The relationship between the FIR luminosity and the SFR is
complex, since stars with a variety of ages can contribute to the dust heating, and only
a fraction of the bolometric luminosity of the young stellar population is absorbed by dust
(e.g., see \citealt{1987ApJ...314..513L,1996ApJ...460..696W}). 
 By adopting the mean luminosity for
10-100 Myr continuous bursts, solar abundances, the \citet{1955ApJ...121..161S} initial mass function (IMF) and assuming that the dust reradiates all of the bolometric luminosity yields:
\begin{equation}\label{eq:sfr}
\rm{\Bigg(\frac{SFR_{IR}}{M_{\odot}\,yr^{-1}}\Bigg)\,=\,\Bigg(\frac{L_{IR}}{5.8\times10^{9}\,L_{\odot}}\Bigg)} \end{equation}
(see \citealt{1998ApJ...498..541K,2003ApJ...586..794B,2011ApJ...737...67M}).
To compute the SFRs, we use the redshift dependent $\rm{q_{IR}(z)}$ parameter. This should account for these intrinsic observational limitations under the assumption of a linear IR-radio correlation given by:
\begin{equation}\label{eq:sfr_dens}
\rm{\Bigg(\frac{SFR_{610\,MHz(z)}}{M_{\odot}\,yr^{-1}}\Bigg)\,=\,\mathcal{F}\,_{IMF}\times\,10^{-24}\,10^{q_{IR}(z)}\Bigg(\frac{L_{610\,MHz}}{WHz^{-1}}\Bigg)} \end{equation}
where $\rm{\mathcal{F}\,_{IMF}}$ = 1 for a \citet{2003PASP..115..763C} IMF and $\rm{\mathcal{F}\,_{IMF}}$ = 1.7 for a \citet{1955ApJ...121..161S} IMF.
\citet{2017A&A...602A...5N} stresses that since low-mass stars do not contribute significantly
to the total light of the galaxy, only the mass-to-light
ratio is changed when the IMF adopted is \citet{2003PASP..115..763C}.  We followed 
\citet{2017A&A...602A...5N} and used the \citet{2003PASP..115..763C} IMF.

\begin{figure}
\centering
\centerline{\includegraphics[width = 0.55\textwidth]{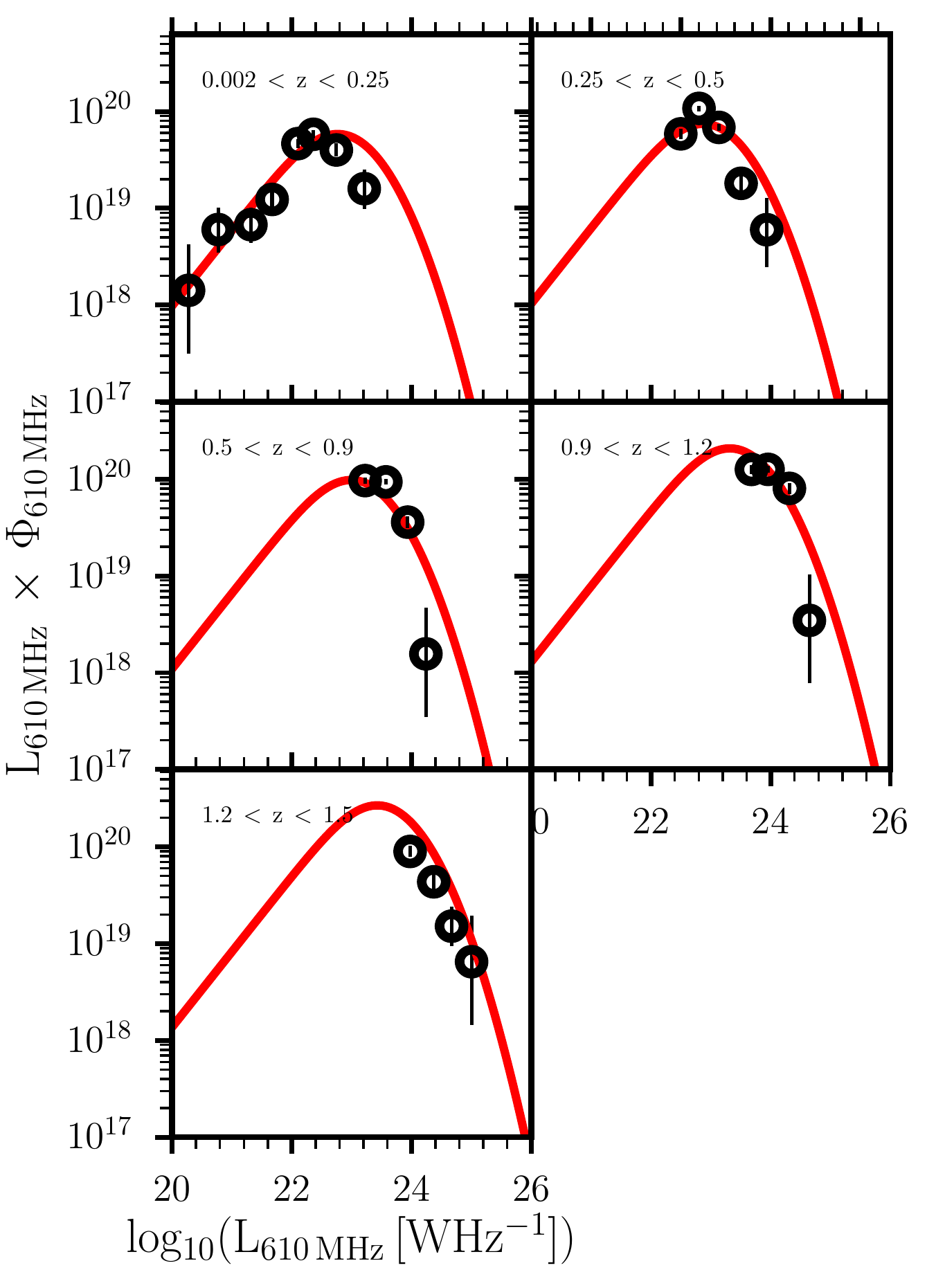}}
\caption{Luminosity density for 610 MHz GMRT star forming galaxies (open black circles) in 5 redshift bins. The solid red  curves correspond to the best fit PLE  LFs in each redshift bin.}
\label{lumdens.fig} 
\end{figure}

\begin{figure}
\centering
\centerline{\includegraphics[width = 0.5\textwidth]{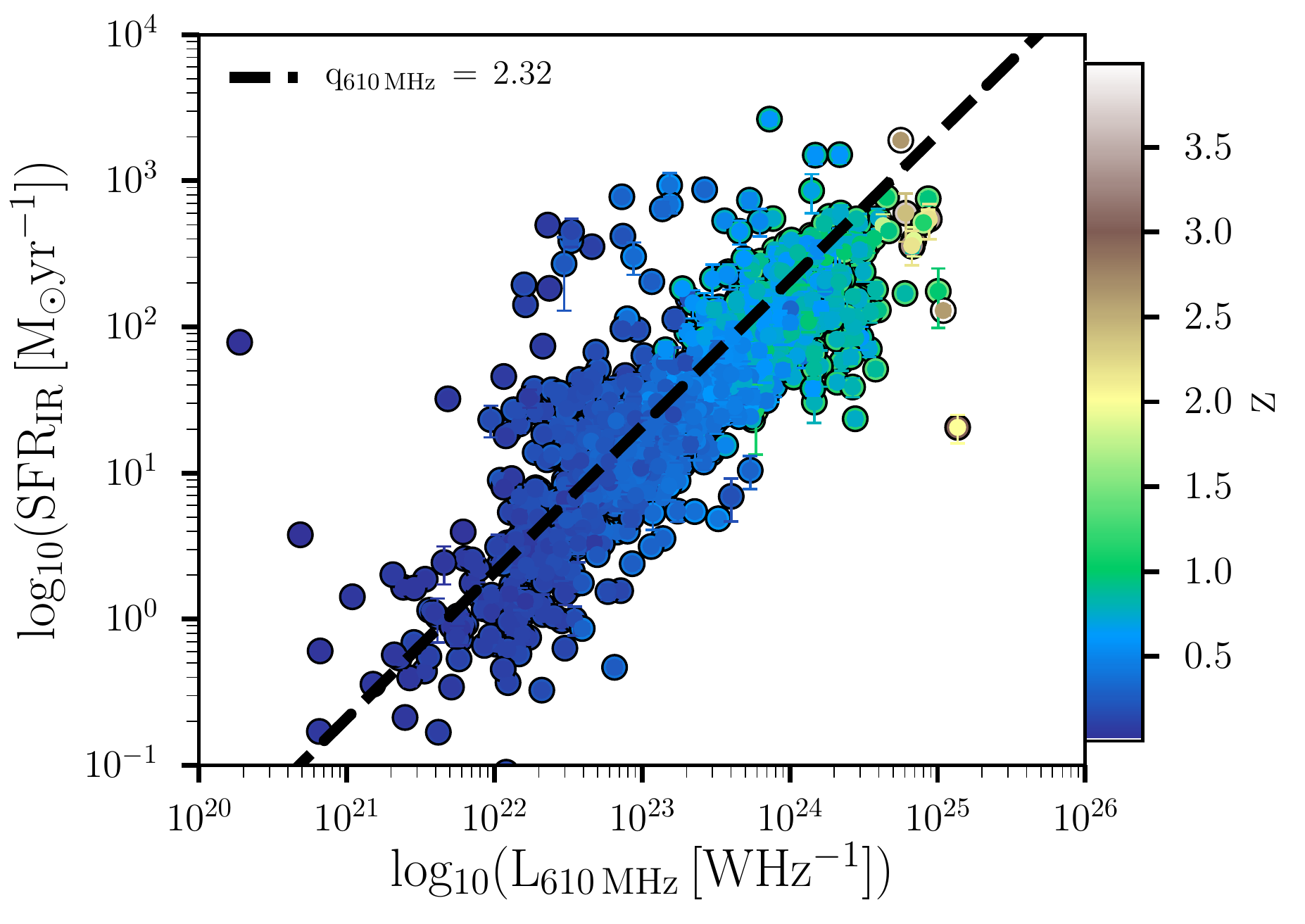}}
\caption{SFR  from the total infrared luminosity as a function of radio luminosity at 610 MHz for SFGs. The SFGs are color coded with redshift. The dotted line shows the SFR, when a non-evolving q-value (i.e.  median q-value at 610 MHz in Section~\ref{IRRC}) is assumed.}
\label{sfr.fig} 
\end{figure}

\begin{figure*}
\centering
\centerline{\includegraphics[width = 0.7\textwidth]{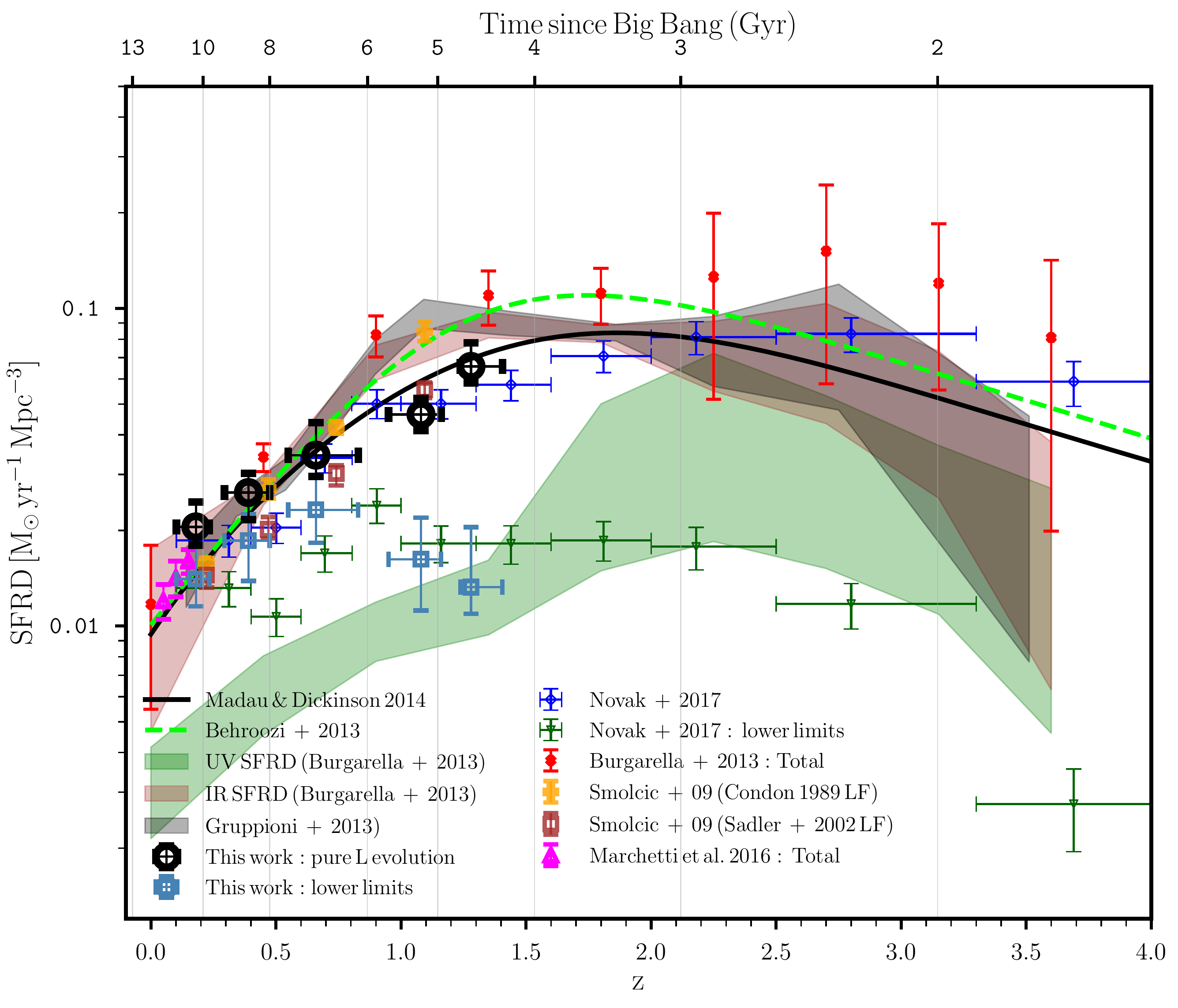}}
\caption{Cosmic star formation rate density  history. Our total SFRD values estimated from the pure luminosity evolution in separate redshift bins are shown as open black circles. See text for details of the
data points shown for comparison that are indicated in the legend. We limit the plot between $\rm{0<z<4}$ redshift range.}
\label{sfr_hist.fig} 
\end{figure*}

In Figure~\ref{sfr.fig} we show SFR  from the total infrared luminosity as a function of radio luminosity at 610 MHz for SFGs. We color code the SFGs with redshift and dotted line shows the SFR, when a non-evolving q-value (i.e. median q-value in Section~\ref{IRRC}) is assumed.
Converting the radio luminosity to SFR as shown in equation~\ref{eq:sfr} to equation~\ref{eq:sfr_dens}, before performing the 
integration will yield the star formation rate density (SFRD) of a given epoch as shown in equation~\ref{eq:sfrd} below, as presented as well by \citet{2017A&A...602A...5N}.
\begin{equation}\label{eq:sfrd}
\rm{\mathcal{SFRD}\,=\,\int_{L_{min}}^{L_{max}}\,\Phi_{(L,z,k_{L},k_{D})}\,\times\,SFR(L)\,d(\log\,L_{610\,MHz})}
\end{equation}
To derive the SFR density we need to compute the  radio luminosity density and to convert our radio luminosities into SFRs, in Figure~\ref{lumdens.fig}, we show the luminosity density for our 5 redshift bins. The curves are the PLE (solid red) best fit to the 610 MHz data in each redshift bin. We numerically integrated the expression in equation~\ref{eq:sfrd} by taking the analytical form of the LF in each redshift bin and using the best fit evolution parameters shown in Figure~\ref{lf_610_sfg}. We integrated over the entire luminosity range by setting $\rm{L_{min}\,=\,0}$ and $\rm{L_{max}\,=\,+\infty}$. This ensures that the integral converges and that the major contribution to the SFRD arises from galaxies with luminosities around the turnover of the LF. From this approach, \citet{2017A&A...602A...5N} stresses that the entire radio emission is recovered and if the LF shape and evolution is well constrained, the SFRD estimate will be within the SFR calibration errors. We also performed the integration using the data constrained limits, where $\rm{L_{min}}$ and $\rm{L_{max}}$ correspond to the lowest and the highest value of the observed LF. This ensures that, any bias due to LF extrapolation toward higher or lower luminosities is removed (see \citet{2017A&A...602A...5N}). We show our total SFRD derived by integrating the pure luminosity evolved LF in individual redshift bins as open black circles in Figure~\ref{sfr_hist.fig}. Table~\ref{evo.tab} presents the  best-fit evolution parameters obtained by the fitting local
luminosity function to the redshift binned data assuming pure luminosity $\rm{k_{L}}$ evolution and the SFRD derived.
We compare our SFRD results with other radio-based estimates in Figure~\ref{sfr_hist.fig}.  \citet{2009ApJ...690..610S} derived the cosmic star formation history out to $\rm{z\,=\,1.3}$ using the local 20 cm LFs (\citealt{1989ApJ...338...13C, 2002MNRAS.329..227S}), purely evolved in luminosity,
and best fit to the VLA-COSMOS data in four redshift bins (see yellow pluses and brown squares). SFRD obtained when $\rm{L_{min}}$ and $\rm{L_{max}}$ are data constrained limits (lower limits) are also shown (see lightblue squares). Lower limits obtained by \citet{2017A&A...602A...5N} are shown as green triangles. To create a consistent multi-wavelength
picture, we also compare our work with results in the literature derived at
 infrared (IR) and ultraviolet (UV) wavelengths in Figure~\ref{sfr_hist.fig}. All SFR estimates were rescaled to a Chabrier IMF where necessary. 
 The curve from the review by \citet{2014ARA&A..52..415M},
who performed a fit on a collection of previously published  UV and IR SFRD data is shown as a solid black curve. The curve from \citet{Behroozi_2013} who provide new fitting formulae for star formation histories based on a wide variety of observations is shown as dashed green curve is also shown. The constained SFRD  by \citet{2013A&A...554A..70B} taking into account dust obscuration using
combined IR and UV LFs reported in \citet{2013MNRAS.432...23G} and \citet{2012A&A...539A..31C}, respectively are shown as red crosses for the total SFRD, red shaded area for the IR SFRD and green shaded area for the UV SFRD. The grey shaded area denote the $\rm{1\sigma}$ uncertainty
for the SFRD derived from integrated total IR LF by \citet{2013MNRAS.432...23G}. SFRD estimates
including the unobscured contribution
based on the UV dust-uncorrected emission from local galaxies by \citet{10.1093/mnras/stv2717} are shown as magenta triangles.
The expected steep decline in the star formation rate density since $\rm{z\sim1}$ seen by \citet{2009ApJ...690..610S} and other previous studies is reproduced by our sample.

\begin{table}
 \centering
 \caption{Best-fit evolution parameters obtained by the fitting local luminosity function to the redshift binned data assuming pure luminosity $\rm{k_{L}}$ evolution and the star formation rate density derived.}
 \begin{tabular}{|c|c|c|c|}
 \hline
 \hline
Med(z)    & $\rm{k_{L}}$ &Total&Lower limits \\
& &$\rm{SFRD\,[M_{\odot}\,yr^{-1}\,Mpc^{-3}]}$&\\
 \hline
 0.18$_{-0.07}^{+0.05}$&2.88$\pm0.48$&0.021$_{-0.003}^{+0.004}$&0.014$_{0.002}^{0.004}$ \\ 
 0.39$_{-0.10}^{+0.09}$& 2.89$\pm0.49$&0.026$_{-0.005}^{+0.004}$&0.018$_{0.004}^{0.003}$ \\
 0.66$_{-0.11}^{+0.16}$& 2.77$\pm0.46$&0.035$_{-0.005}^{+0.009}$&0.023$_{0.005}^{0.009}$\\
 1.08$_{-0.13}^{+0.08}$& 2.67$\pm0.41$&0.046$_{-0.005}^{+0.006}$&0.016$_{0.005}^{0.006}$\\ 
 1.28$_{-0.04}^{+0.13}$& 2.57$\pm0.40$&0.066$_{-0.007}^{+0.012}$&$0.013_{0.002}^{0.007}$\\ 
 
\hline
\end{tabular}
\label{evo.tab} 
\end{table}

\begin{table*}
 \centering
 \caption{Luminosity functions of SFG obtained with the $\rm{1/V_{max}}$ method for different redshift bins.}
 \begin{tabular}{|c|c|c|c|}
 \hline
 \hline
Redshift & Luminosity    & Number density & Number \\
$\rm{z}$&$\mathrm{\log_{10}(L_{610\,MHz}\,[WHz^{-1}])}$&$\rm{\Phi_{610}(Mpc^{-3}dex^{-1})}$ & N\\
 \hline
 $\rm{0.002\,<\,z\,<0.25}$&20.28&7.51$_{-5.70}^{+14.64}$ $\rm{\times\,10^{-3}}$    & 1     \\
 &20.77& 1.02$_{-5.70}^{+14.64}$ $\rm{\times\,10^{-2}}$   & 4\\
  &21.32& 3.25$_{-5.70}^{+14.64}$ $\rm{\times\,10^{-3}}$   & 6\\
  &21.67& 2.63$_{-5.70}^{+14.64}$ $\rm{\times\,10^{-3}}$   & 20\\
   &22.10&3.68$_{-5.70}^{+14.64}$ $\rm{\times\,10^{-3}}$   & 68\\
&22.36&2.53$_{-5.70}^{+14.64}$ $\rm{\times\,10^{-3}}$   & 89\\
 &22.74&7.21$_{-5.70}^{+14.64}$ $\rm{\times\,10^{-4}}$   & 36\\ 
&23.21&9.81$_{-5.70}^{+14.64}$ $\rm{\times\,10^{-5}}$   & 5\\
 \hline   
  $\rm{0.25\,<\,z\,<0.5}$&22.50&1.86$_{-0.21}^{+0.24}$ $\rm{\times\,10^{-3}}$    & 58     \\
  &22.80&1.69$_{-0.11}^{+0.16}$ $\rm{\times\,10^{-3}}$   & 176\\
  &23.14& 5.01$_{-0.40}^{+0.44}$ $\rm{\times\,10^{-4}}$   & 113\\
  &23.51& 5.63$_{-1.21}^{+1.53}$ $\rm{\times\,10^{-5}}$   & 16\\
   &23.93&6.98$_{-4.10}^{+7.95}$ $\rm{\times\,10^{-6}}$   & 2\\
 \hline   
$\rm{0.5\,<\,z\,<0.9}$&23.23&5.75$_{-0.42}^{+0.46}$ $\rm{\times\,10^{-4}}$    & 135     \\
 &23.57& 2.51$_{-0.17}^{+0.18}$ $\rm{\times\,10^{-4}}$   & 163\\
  &23.94& 4.17$_{-0.54}^{+0.63}$ $\rm{\times\,10^{-5}}$   & 43\\
  &24.24& 8.99$_{-6.98}^{+17.93}$ $\rm{\times\,10^{-7}}$   & 1\\
 \hline   
$\rm{0.9\,<\,z\,<1.2}$&23.68&2.64$_{-0.27}^{+0.30}$ $\rm{\times\,10^{-4}}$    & 71     \\
 &23.96& 1.39$_{-0.11}^{+0.12}$ $\rm{\times\,10^{-4}}$   & 117\\
  &24.31& 3.89$_{-0.47}^{+0.54}$ $\rm{\times\,10^{-5}}$   & 49\\
  &24.65& 7.81$_{-6.07}^{+0.16}$ $\rm{\times\,10^{-7}}$   & 1\\
 \hline   
$\rm{1.2\,<\,z\,<1.5}$&23.98&9.46$_{-1.18}^{+1.35}$ $\rm{\times\,10^{-5}}$    & 47     \\
 &24.37& 1.85$_{-0.32}^{+0.39}$ $\rm{\times\,10^{-5}}$   & 24\\
  &24.67& 3.25$_{-1.24}^{+1.89}$ $\rm{\times\,10^{-6}}$   & 5\\
  &25.00& 6.44$_{-5.00}^{+12.84}$ $\rm{\times\,10^{-7}}$   & 2\\
     
\hline
\end{tabular}
\begin{tablenotes}
\item [a] The listed luminosity values represent the median luminosity of the sources in the corresponding luminosity bin.
\end{tablenotes}
\label{LF_bins.tab} 
\end{table*} 

\section{Conclusions}\label{sum.sec}
 Over the last few years \citet{2013MNRAS.436.3759B}, \citet{2015MNRAS.452.1263P} and \citet{2017A&A...602A...1S} have for the first time managed to carry out a complete census of populations contributing to the faint radio sky at 1.4 and 3.0 GHz with the JVLA. These were done over a relatively small (0.5 deg$^2$ and 2 deg$^2$ for JVLA-ECDFS and JVLA-COSMOS respectively) contiguous area reaching rms sensitivities (i.e. 6$\mu$Jy and 2.3$\mu$Jy at 1.4 and 3.0 GHz respectively) comparable to our study. Such datasets that provide images of the radio flux density over small regions at these sensitivities are still rare but will be achieved by the MeerKAT and SKA1 albeit over much larger areas.
We  study a sample of 1685 SFGs covering $\sim$1.86 deg$^2$ down to a a minimum noise of $\sim$7.1\,$\mu$Jy / beam in the EN1 field at 610 MHz with the GMRT. The depth of our 610 MHz data represent a potentially very useful tool to address the role of SFGs in galaxy evolution. These SFGs were obtained from a combination of diagnostics from the radio and X-ray luminosity, optical spectroscopy, mid-infrared colors, and 24$\mu$m and IR to radio flux ratios. Of the 1685 SFGs from our sample, 496 have spectroscopic redshifts whereas 1189 have photometric redshifts.
 Deep multi-wavelength spectrophotometric datasets with comparable resolutions and sensitivities to our radio data will be needed to improve our source classification. More specifically, mid-infrared multi-band photometry and optical/near-infrared spectroscopy are the limiting factor in our case. However, since no deep (wide-area) mid-infrared will not be carried out in the foreseeable future, the best avenue toward improving the diagnostics for such scientific work in the future is via multi-object wide-field optical/near-infrared spectroscopy.

We study the infrared-radio correlation (IRRC) for the star-forming galaxies. We measure an evolution with redshift of the IRRC for 1.4 GHz radio luminosities to be $\rm{q_{IR}\,=\,2.86\pm0.04(1\,+\,z)^{-0.20\pm0.02}}$, where $\rm{q_{IR}}$ is the ratio between the total
infrared luminosity $\rm{(L_{IR}, 8-1000\mu\,m)}$ the 1.4 GHz radio luminosity $\rm{(L_{1.4}\,GHz)}$.

We have used the non-parametric V/Vmax test and the radio luminosity function to investigate the cosmic evolution of SFGs. \cite{2007MNRAS.381..211S} found evidence that low-luminosity radio sources experience mild evolution with an increase in their number density by a factor of $\rm{\sim2}$ at $\rm{z\,=\,0.55}$.
\cite{2009ApJ...690..610S} found a mild evolution of the low-power AGNs in the VLA-COSMOS survey out to $\rm{z\sim1.3}$. We  construct the RLF at 610 MHz for our SFGs and find positive evolution. This is consistent with previous studies, for the SFG RLF   scaled to 610 MHz from 1.4 GHz assuming a spectral index of $\rm{\alpha\,=\,-0.8}$. The exact shape of the radio spectral energy distribution (SED) of SFGs is usually assumed to be a superposition of the steep synchrotron spectrum, described by a power law (see \citealt{1992ARA&A..30..575C}), even so,
there are processes which can alter the shape of the spectra. Recent work by \cite{2013MNRAS.431.3003L} and \cite{2019A&A...621A.139T} have developed theoretical models describing an alternative picture to the simple power-law
shape which includes spectral curvature.
However, deep multi-frequency radio observations of representative samples of galaxies are needed to study the radio SED and understand the physical processes shaping it  across redshifts. We  also compare our results to models from the literature and find that the \citet{Mancuso2017} and \cite{2008MNRAS.388.1335W} models do compare well to our data. However, there is an exception for the lowest redshift bin, where none of the models is able to reproduce the low-luminosity observations. Our LFs behave very well at high luminosities, where other samples (see e.g. \citet{2009ApJ...696...24S}) show an excess of sources with respect to models. This can be interpreted as contamination due to AGN. This can be better addressed with better multi-wavelength data and better-proven AGN diagnostics. Our radio LFs can be well described by a local LF evolved only in luminosity as $\rm{L_{610\,MHz}\,\propto\,(\,1+\,z)^{(2.95\pm0.19)-(0.50\pm0.15)z}}$.

 We converted our radio luminosities to SFRs using a redshift dependent IR-radio correlation.
 By integrating over the entire luminosity range the LF fits in various redshift bins, we derived the cosmic star formation density out to $z = 1.5$ for our SFG sample. Our estimates of the SFRD is 
consistent with previous measurements from the literature when all the SFR estimates are rescaled to a Chabrier IMF. \citet{2017A&A...602A...5N} assumed pure luminosity evolution for their LF, consistent
with the measurements by \citet{2009ApJ...696...24S} (and recent result by \citet{2013MNRAS.432...23G} assuming the redshift dependent IR-radio correlation parameter). All these studies found broad agreement between the radio SFRD evolution and UV/IR surveys, observing a steep decline from $z = 1$ to 0. Our sample reproduces this expected steep decline in the star formation rate density since $\rm{z\sim1}$. This work represents a  benchmark for studying the evolution of the RLF and SFR function with
cosmic time at the faint low-frequency regime in spite of our redshift limit.

In the near future we plan to undertake the exploitation of the MeerKAT International GHz Tiered Extragalactic Exploration (MIGHTEE) Survey \citep{MIGHTEE2016} with the MeerkAT SKA precursor \citep{MEERKAT2016}. MIGHTEE  will survey well-studied extragalactic deep fields (E-CDFS, COSMOS, XMM-LSS and ELAIS-S1), totaling 20 square degrees at 1.4 GHz, to $\sim$2\,$\mu$Jy/beam rms.
A survey matched in resolution and depth will be undertaken with the upgraded GMRT \citep{Gupta2014}, and the present work will thus be precious to make the most of such GMRT data. It is our  hope that a complete and expansive review of this topic will be able to do justice to
the wealth of current and ongoing measurements contributing to our understanding of this aspect of galaxy evolution. An extensive compilation from the literature of SFR density measurements as a function of redshift will be investigated in future works. This will provide rich compilation of SFR density evolution which will help with a robust constraint for many investigations of galaxy evolution.

\appendix

\section{LF}
\begin{figure}
\centering
\centerline{\includegraphics[width = 0.55\textwidth]{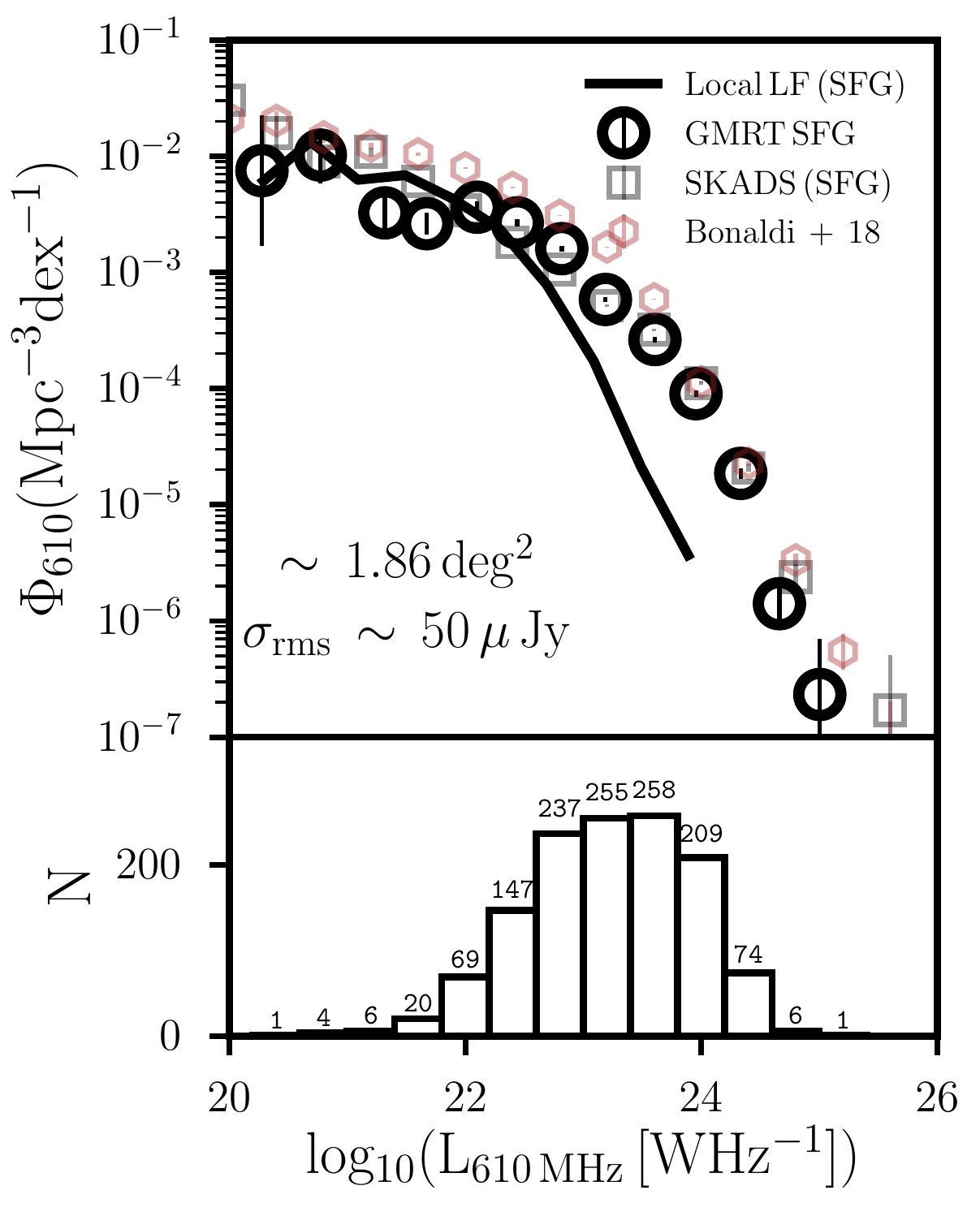}}
\caption{The 610 MHz RLF for SFGs in the redshift range $\rm{0.002\, < \,z \,<\,1.5}$ shown in open black circles. The local RLF of \citet{2007MNRAS.375..931M}, which has been converted to 610 MHz from 1.4 GHz assuming a spectral index of $\rm{\alpha\,=\, -0.8}$, is shown in black line.
The LF's computed for SFGs from T-RECS \citep{Bonaldi2019} simulations are shown as open brown diamonds.
Error bars are determined using the
prescription of \citet{1986ApJ...303..336G}.
The lower panel shows the luminosity distribution with the number of sources in each bin indicated on the bar.
}
\label{lumfunc_all.fig} 
\end{figure}

The LFs obtained from the $\rm{\frac{1}{V_{max}}}$ method for SFGs for the entire redshift range we consider (i.e. $\rm{0.002\, < \,z \,<\,
1.5}$) is shown in Figure~\ref{lumfunc_all.fig} as open black circles. The local RLF from \cite{2007MNRAS.375..931M} is represented as black solid line in this figure for comparison. The total SFG RLF for our sample is higher than the SFG RLF of \cite{2007MNRAS.375..931M} scaled to 610 MHz assuming $\rm{\alpha\,=\,-0.8}$. This is evident especially at the high luminosity end and can be attributed to cosmic evolution of the SFGs known to positively evolve with redshift \citep{2012MNRAS.426.3334M}. Also, this may be attributed to the sensitivity of our radio observations which allows us to probe the source population up to high redshifts. The open brown diamonds show LF's computed for SFGs from T-RECS \citep{Bonaldi2019} simulations. The faded black squares are RLF computed from the semi-empirical simulation of the SKA \citep{2008MNRAS.388.1335W}. This is in good agreement with the RLF we compute for the SFG sample over the same redshift range.
The breakdown of the results obtained for the 610 MHz RLF for SFGs from $\rm{0.002\, < \,z \,<\,1.5}$ is presented in Table~\ref{LF_all.tab}.

\begin{table}
 \centering
 \caption{610 MHz RLF for SFGs from $\rm{0.002\, < \,z \,<\,1.5}$.}
 \begin{tabular}{|c|c|c|}
 \hline
 \hline
Luminosity    & Number density & Number \\
$\mathrm{\log_{10}(L_{610\,MHz}\,[WHz^{-1}])}$&$\rm{\Phi_{610}(Mpc^{-3}dex^{-1})}$ &\\
 \hline
 20.28&7.51$_{-5.70}^{+14.64}$ $\rm{\times\,10^{-3}}$    & 1     \\
 20.77& 1.02$_{-3.34}^{+6.67}$ $\rm{\times\,10^{-2}}$   & 4\\
 21.32& 3.25$_{-1.52}^{+2.61}$ $\rm{\times\,10^{-3}}$   & 6\\
 21.67& 2.63$_{-0.54}^{+0.50}$  $\rm{\times\,10^{-3}}$   & 20\\
 22.09& 3.64$_{-0.24}^{+0.28}$  $\rm{\times\,10^{-3}}$  & 69\\
 22.44& 2.67$_{-0.15}^{+0.17}$  $\rm{\times\,10^{-3}}$   & 147\\
 22.82& 1.60$_{-0.63}^{+0.68}$ $\rm{\times\,10^{-3}}$   & 237\\
 23.19& 5.85$_{-0.25}^{+0.27}$ $\rm{\times\,10^{-4}}$  & 255\\
 23.61& 2.64$_{-0.09}^{+0.09}$ $\rm{\times\,10^{-4}}$   & 258\\
 23.95& 9.04$_{-0.39}^{+0.42}$ $\rm{\times\,10^{-5}}$   & 209\\
 24.33& 1.86$_{-0.16}^{+0.18}$ $\rm{\times\,10^{-5}}$   & 74\\
 24.66& 1.41$_{-2.63}^{+5.10}$ $\rm{\times\,10^{-6}}$   & 6\\
 25.00& 2.33$_{-2.63}^{+5.10}$ $\rm{\times\,10^{-7}}$   & 1\\
\hline
\end{tabular}
\begin{tablenotes}
\item [a] The listed luminosity values represent the median luminosity of the sources in the corresponding luminosity bin.
\end{tablenotes}
\label{LF_all.tab} 
\end{table}

\section*{Acknowledgements}

We thank the staff of the GMRT that made these observations possible. GMRT is run by the National Centre for Radio Astrophysics of the Tata Institute of Fundamental Research. We also thank Anna Bonaldi for providing us with her model predictions. This work is based in part on observations made with the \textit{Spitzer Space Telescope}, which is operated by the Jet Propulsion Laboratory, California Institute of Technology under a contract with NASA.
We acknowledge support from the Italian Ministry of Foreign Affairs and International Cooperation (MAECI Grant Number ZA18GR02) and the South African Department of Science and Technology's National Research Foundation (DST-NRF Grant Number 113121) as part of the ISARP RADIOSKY2020 Joint Research Scheme.
\bibliographystyle{mnras}
\bibliography{main} 
%
\bsp	
\label{lastpage}
\end{document}